\documentclass[11pt]{article}

\usepackage{a4wide}
\usepackage{amssymb}
\usepackage{amsmath}
\usepackage{color}
\usepackage{here}
\usepackage{graphics,epsfig}

\newcommand{\nco}{\newcommand}
\newfont{\msbm}{msbm10 at 12pt}
\nco{\CC}{\mbox{\msbm C}}
\nco{\ZZ}{\mbox{\msbm Z}}
\nco{\NN}{\mbox{\msbm N}}
\nco{\II}{\mbox{\msbm I}}
\nco{\RR}{\mbox{\msbm R}}
\nco{\munite}{\ensuremath{\,\,\mathrm{l}\!\!\!1}}
\nco{\dps}{\displaystyle}
\nco{\ov}{\overline}
\nco{\ud}{\underline}

\def\otimesdot{\dot{\otimes}}

\def\ie{{\rm i.e.,\/}\ }
\def\etc{{\rm etc.\/}\ }

\title{Comments about quantum symmetries of $SU(3)$ graphs \vspace{1.0cm}}

\author{
{\bf R. Coquereaux}${}^{1}\;$\thanks{E-mail:Robert.Coquereaux@cpt.univ-mrs.fr},
{\bf D. Hammaoui}${}^{2}\;$\thanks{E-mail:d.hammaoui@sciences.univ-oujda.ac.ma}, {\bf G. Schieber}${}^{1,2,3}$\thanks{Supported by a fellowship of AUF - Agence Universitaire de la Francophonie. E-mail: schieber@cbpf.br}, {\bf E. H. Tahri}${}^{2}\;$\thanks{E-mail:tahrie@sciences.univ-oujda.ac.ma}.  \\
\\
${}^1$ {\it CPT - Centre de Physique Th\'eorique - CNRS} \\
   {\it Campus de Luminy - Case 907}\\
   {\it F-13288 Marseille - France}\\
\\
${}^2$ {\it LPTP - Laboratoire de Physique Th\'eorique et des Particules} \\
{\it D\'epartement de Physique, Facult\'e des Sciences} \\
   {\it Universit\'e Mohamed I, B.P.524 Oujda 60000 - Maroc} \\
\\
${}^3$ {\it CBPF - Centro Brasileiro de Pesquisas F\'{\i}sicas} \\
                 {\it Rua Dr. Xavier Sigaud, 150}\\
                 {\it 22290-180, Rio de Janeiro, Brasil}\\
}

\date{}
\begin{document}
\thispagestyle{empty}

\begin{titlepage}

\maketitle


\begin{abstract}
For  the  $SU(3)$ system of  graphs generalizing the ADE Dynkin
digrams in the classification of modular invariant partition functions in CFT, we present a general 
collection of algebraic objects
and relations that describe  fusion properties and quantum
symmetries  associated with the corresponding Ocneanu quantum
groupo\"{\i}ds. We also summarize the properties of the individual
members of this system.
\end{abstract}

\vfill


\vspace{1.0cm}



\end{titlepage}


\section{Introduction}

\paragraph{The stage}
Along the last fifteen years or so, investigations performed in a
number of research fields belonging to theoretical physics or to
mathematics suggest the existence of  ``fundamental objects''
generalizing the usual simply laced ADE Dynkin diagrams. Let us
mention a few of these fields:  statistical mechanics, string
theory, quantum gravity, conformal field theory, theory of
bimodules, Von Neumann algebras, sector theory, (weak) Hopf
algebras, modular categories, \etc

Properties of the  algebraic structures associated with  the choice of such a fundamental object have been analysed independently by  several groups of people, with their own tools and terminology.
The  results obtained by  these different schools are not always easy to compare, or even to  aprehend,
because of the required background and specificity of the language.

However, at the heart of any such  fundamental object  we meet a
graph (or the adjacency matrix that encodes this graph). We believe that
many important and useful results can be described in an
elementary way obtained from the combinatorial data provided by
the graph itself, or by some kind of attached modular data \cite{ganon-moddata}.

Roughly speaking, if we have a modular invariant (but not any kind
of modular invariant), we have a (particular type of) quantum
groupo\"{\i}d, and conversely. Now every such quantum
groupo\"{\i}d is encoded by a graph, and this graph leads
naturally to two (in general distinct) character theories: one is
the so called fusion algebra, and the other is the algebra of
quantum symmetries. This is the story  that we want to tell. But
we want to tell it in simple words, using elementary mathematics.
And we want to tell it in the case of the $SU(3)$ system of
graphs, \ie the so-called ``Di Francesco - Zuber diagrams'' that
generalize the familiar ADE Dynkin diagrams.

As already mentioned, several groups of people (without trying to be exhaustive, we can cite 
\cite{pasquier-ADE,ganon-moddata,Fuchs,Pet_Zub-Occells,bek,GilRobert1,Robert_Marina,Chui,ostrik,Cal_Etingof}) have
investigated related topics along the past years. We believe
that only A. Ocneanu has actually worked out all these examples in
details, with his own language,  from the point of view of the
study of quantum symmetries, but his results are unfortunately
unpublished and not available.

\paragraph{Purpose}
The purpose of this article is three-fold.

1) To present, in a synthetic and elementary way, a collection of
algebraic objects describing
fusion properties and quantum symmetries associated with graphs
belonging to (higher) Coxeter-Dynkin systems.

2) To present a summary of results concerning members of the $SU(3)$ system.

3) To make a number of comments about the various aspects of this subject, and, in some cases,
to establish a distinction between what  is known and  what is  believed to be true.

\paragraph{Warning}
This paper is not a review. If it is true that many results recalled here can be found in the litterature, maybe with another language or perspective, many others cannot be found elsewhere. It may well be that a  number of these results have been privately worked out by several people, but, if so, they are not available. What we present here, including a good part of the terminology itself, is mostly the result of our own understanding, that has been growing up  along the years.

 However, this paper is not a detailed research paper either. Indeed, it is, in a sense, too short. Every single example summarized in section $6$, for instance, gives rise to interesting, and, sometimes difficult, problems, and would certainly be worth a dedicated article. What we put in this section is only what we think should be remembered once all the details will have been forgotten. This, admitedly, is a partial viewpoint.

We want this paper to be used as a compendium of results,
terminology, and remarks.

\paragraph{Plan}
 The plan of this article is as follows.
 In the next section we summarize the properties of the ${\mathcal A}$ system, \ie the Weyl alcoves at level $k$, from the viewpoint of fusion and graph algebras.
 In section $3$, we describe  general properties associated with any member of  the $SU(3)$ system of graphs. This applies, in particular, to the ${\mathcal A}$ graphs themselves, but they are very particular, and this is why we singled them out.
 In the fourth section, we describe, in plain terms, the Ocneanu quantum groupo\"{\i}d associated with a graph $G$, or, better, with a pair $(G,{\mathcal A}_k)$. We do not give however any information about the methods that allow one to compute the values of the corresponding cells; this is a most essential question but it should be dealt with in another publication.  In the fifth section we describe the equations that allow one to recover the algebra of quantum symmetries (and sometimes the graph itself) from the data provided by a modular invariant, the leitmotiv of this section being the so-called ``modular splitting technique''.  Although we have used repeatedly this technique to solve several quite involved examples briefly described in section $6$, we do not explicitly discuss here our method of resolution  but refer to forthcoming articles (or theses)  for these -- important -- details \cite{EstebanGil, GilDahmaneHassan2, Dahmane-thesis}.
In section $6$ we summarize what is known, or at least what we
know, about the structure of the algebra of quantum symmetries for
each member of the $SU(3)$ series. At this point we should stress
that the graphs themselves, together with their fusion properties
(relations with the ${\mathcal A}$ system) or with the associated
modular invariants, have been discovered and described long ago
(by Di Francesco and Zuber \cite{DiFZuber}). Several aspects
related to the theory of sectors, or to the theory of bimodules
have also been investigated independently by different groups of
people \cite{be,bek,Fuchs,Evans}. However we believe that only A. Ocneanu performed a
detailed analysis of the algebra of quantum symmetries associated
with all these diagrams and three of us remember vividly the poster
describing the Cayley graph for the generators of the algebra that
we call $Oc({\mathcal E}_9)$,  on one of the walls of the
Bariloche conference lecture hall, during the January 2000 summer
(!) school. However, this material was never published or even
made public on the internet. Our techniques may be sometimes
clumsy but we hope that they are understandable and will draw
attention of potential readers on this fascinating subject.  We
now return to the plan of our paper and mention that the last
section (the 7th) is devoted to a set of  final remarks describing
possible new directions or open problems.

\section{${\cal A}_k$ graphs}

\subsection{First properties}
The ${\cal A}_k$ graphs are obtained as truncations of the Weyl chambers of
$SU(N)$ at some level (Weyl alcoves). They have a level $k$
and a (generalized) Coxeter number $\kappa =  k+N$. From now on $N=3$.

\paragraph{Vertices}
Vertices $\lambda$ may be labelled by Dynkin labels $(\lambda_1, \lambda_2)$,
with $0 \leq \lambda_1 + \lambda_2 \leq k$, by shifted Dynkin labels
$\{\lambda_1 +1, \lambda_2 +1\}=(\lambda_1, \lambda_2)$,
or by Young tableaux\footnote{$p$ (resp. $q$) is the number of boxes in the first (resp. second) line.}
$Y[p,q]$, $p=\lambda_1+\lambda_2$, $q=\lambda_2$. For instance, the unit vertex (trivial representation) is
$(0,0) = \{1,1\} = Y[0,0]$, the fundamental vertex $(1,0) = \{2,1\} = Y[1,0]$
and its conjugate $(0,1) = \{1,2\} = Y[1,1]$.
The graph ${\cal A}_k$ possesses $d_{{\cal A}_k} = (k+1)(k+2)/2$ vertices.
The vector space spanned by these vertices is also called
${\cal A}_k$.

\paragraph{Conjugation}
The graph ${\cal A}_k$ has an involution $\star$:
$(\lambda_1,\lambda_2) \rightarrow (\lambda_2,\lambda_1)$ called conjugation.

\paragraph{Triality}
Each vertex $\lambda$ possesses a triality $t(\lambda)=\lambda_1-\lambda_2 \mod 3$.
It is equal to the number of boxes modulo $3$ of the corresponding Young tableau. Conjugation leaves triality 0 invariant and interchanges 1 and 2. 

\paragraph{Edges}
Edges are oriented. They only connect vertices of increasing triality, by step +1, i.e. we choose one of the two possible adjacency matrices (the other is its transpose, with the edges in the opposite direction). 

\subsection{Spectral properties}

\paragraph{Exponents and norm}
The adjacency matrix of the graph ${\cal A}_k$ possesses $d_{{\cal A}_k}$ distinct complex
eigenvalues\cite{Zub-fusion}:
\begin{equation}
\beta(r_1,r_2) = e^{-\frac{2 i \pi (2(r_1+1)+(r_2+1))}{3 \kappa}} \left(1+e^{\frac{2 i \pi(r_1+1)}\kappa}
+ e^{\frac{2 i \pi \left( (r_1+1)+(r_2+1) \right) }{\kappa}} \right) \;,
\label{valprop}
\end{equation}
where $r_1,r_2 \geq 0$ and $r_1+r_2 \leq k$. Such pairs of
integers $(r_1,r_2)$ are called exponents of the graph ${\cal
A}_k$. The vertices of the ${\cal A}_k$ graph can be indexed by
the same set of integer pairs $(r_1,r_2)$: they coincide with the
Dynkin labels $(\lambda_1,\lambda_2)$. The set of eigenvalues is
invariant under the group $\mathbb{Z}_3$. One of these eigenvalues
$\beta \doteq \beta(0,0)$ is real, positive, and of largest
absolute value. It is called the norm of the graph, and it is
equal to $\beta = 1 + 2 \, cos(2 \pi/\kappa)$.

\paragraph{Class vectors, dimension vector and quantum dimensions}
Normalized eigenvectors of the adjacency matrix are
denoted $c_{r_1,r_2}$. They can be called ``class vectors'' in analogy with the
situation that prevails for finite groups. Here ``normalized'' means that the first
component\footnote{We assume that an order has been chosen on the set of vertices and that the unit vertex comes first.} of each class vector, corresponding to the unit vertex, is set to $1$. The normalized eigenvector associated with the biggest eigenvalue $\beta$ is called the dimension vector, or the Perron-Frobenius vector. Its components define the quantum
dimensions of the corresponding vertices of ${\cal A}_k$. The quantum dimension of a given vertex $\lambda=(\lambda_1,\lambda_2)$ is given by the q-analog of the classical formula for dimensions of $SU(3)$ irreps, usual numbers being replaced by quantum numbers: $qdim(\lambda)=(1/[2]_q)([\lambda_1+1]_q[\lambda_2+1]_q[\lambda_1+\lambda_2+2]_q)$,
where $q=\exp ( i\pi /\kappa) $ is a root of unity and
$\left[n\right] _{q}=\frac{q^{n}-q^{-n}}{q-q^{-1}}$.  The norm $\beta$
itself is the quantum dimension of the fundamental vertices (1,0) and (0,1).
The sum of $qdim(\lambda)^2$ is called the order or the quantum mass of ${\cal A}_k$ and denoted $m({\cal A}_k)$.

\subsection{Fusion algebra}
The vector space ${\cal A}_k$ possesses an associative (and
commutative) algebra structure: it is an algebra with unity, vertex $(0,0)$, and
two generators, vertices $(1,0)$ and  $(0,1)$,
called ``fundamental generators''. The graph of
multiplication by the first generator $(1,0)$ is encoded by the
(oriented) graph ${\cal A}_k$: the product of a given vertex
$\lambda$ by the fundamental $(1,0)$ is given by the sum of vertices
$\mu$ such that there is an edge going from $\lambda$ to $\mu$ on the graph.
Equivalently, this multiplication is encoded by the
adjacency matrix $N_{(1,0)}$ of the graph.
Multiplication by the other fundamental generator is obtained by
reversing the arrows.

\paragraph{Fusion matrices}
Multiplication by generators $\lambda = (\lambda_1,\lambda_2)$
is described by matrices $N_{\lambda}$, called fusion matrices. The identity is
$N_{(0,0)}=\munite_{d_{\cal A}}$.
The other fusion matrices are obtained, once $N_{(1,0)}$ is known, from the known recurrence relation for
coupling of  irreducible $SU(3)$ representations (that we
-- of course -- truncate at level $k$):
\begin{eqnarray}
N_{(\lambda,\mu)} &=& N_{(1,0)} \, N_{(\lambda-1,\mu)} - N_{(\lambda-1,\mu-1)} -
N_{(\lambda-2,\mu+1)} \qquad \qquad \textrm{if } \mu \not= 0 \nonumber \\
N_{(\lambda,0)} &=& N_{(1,0)} \, N_{(\lambda-1,0)} - N_{(\lambda-2,1)}
\label{recN} \\
N_{(0,\lambda)} &=& (N_{(\lambda,0)})^{tr} \nonumber
\end{eqnarray}
where matrices $N_{(\lambda,\mu)} =0 $ if $\lambda = -1$ or $k+1$ or if $\mu = -1$ or $k+1$, and are periodic in the ($\lambda, \mu$) plane -- the periodicity cell is a Weyl alcove and there are six of them around the origin \{1,1\} = (0,0).
These matrices have non negative integer entries $(N_{\lambda})_{\mu \nu}=N_{\lambda\mu}^{\nu}$
called fusion coefficients. They form a faithfull representation of the fusion algebra:
\begin{equation}
N_{\lambda} N_{\mu} = \sum_{\nu}  N_{\lambda\mu}^{\nu} \, N_{\nu} \;.
\end{equation}
Conjugation (operation $\star$) on these matrices is obtained by transposition.

\paragraph{Essential paths (also called horizontal paths)}
Since fusion matrices $N_{\lambda}$ have non negative integer
entries, one can associate a graph to every fusion matrix.
If the matrix element $(N_{\lambda})_{\mu \nu} = p$, we introduce
$p$ oriented edges from the vertex $\mu$ to the vertex $\nu$.  Such an
edge is called an essential path of type $\lambda$ from $\mu$ to
$\nu$. Remember that these indices are themselves Young tableaux.
The graph associated with the fundamental
generator (1,0) is the ${\cal A}_k$ graph itself.

\subsection{Modular considerations}
The graphs ${\cal A}_k$ support a representation of the group $SL(2,Z)$. This group is generated by two transformations $S$ and $T$ satisfying $S^2=(ST)^3=C$, with $C^2=1$. The modular group itself, called $PSL(2,Z)$ is the quotient of this group by the relation $C = 1$.

\paragraph{The modular generator $S$}
The adjacency matrix of ${\cal A}_k$ can be diagonalized by a matrix constructed from the set of eigenvectors
(all eigenvalues are distinct). As fusion matrices $N_{\lambda}$ commute, this matrix therefore diagonalizes all fusion
matrices. Each line of this matrix is given by a (renormalized)  class vector. We renormalize the lines in order that each line is of norm 1. We therefore divide each class vector by its norm.
The obtained diagonalizing matrix is then unitary but not a priori symmetric, and not necessarily related to the generator of the modular group. To write such an unitarizing matrix, one has first to choose an order on the set of  eigenvalues (this fixes the ordering of line vectors), and also an order on the set of  vertices of the graph (this fixes the ordering of the components for each line). One member of this family of unitarizing matrices gives the modular generator $S$.
The point is that vertices of the graph ${\cal A}_k$ have to be indexed by the same set of integers as the eigenvalues themselves\footnote{We thank O. Ogievetsky for this remark.}.
So, whatever the order we choose on the set of vertices, we decide to choose the {\sl same\/} order on the set of eigenvalues. This procedure determines -- for each ordering of the vertices --  a particular unitarizing matrix which can be identified with the modular generator $S$.
It coincides with the expression explicitly given by the formula \cite{Kac-Peterson, Gannon-su3}:
\begin{eqnarray*}
S_{\lambda \mu } &=&\frac{-i}{\sqrt{3}\kappa }\left( e_{\kappa}
[2\lambda_1\mu_1 + \lambda_1\mu_2 + \lambda_2\mu_1 + 2\lambda_2\mu_2]
-e_{\kappa}[-\lambda_1\mu_1 + \lambda_1\mu_2 + \lambda_2\mu_1 + 2\lambda_2\mu_2]\right. \\
 && \qquad -e_{\kappa}[2\lambda_1\mu_1 + \lambda_1\mu_2 + \lambda_2\mu_1 - \lambda_2\mu_2]
+e_{\kappa}[-\lambda_1\mu_1 + \lambda_1\mu_2 - 2\lambda_2\mu_1 - \lambda_2\mu_2] \\
 && \left. \qquad +e_{\kappa }[-\lambda_1\mu_1 - 2\lambda_1\mu_2 + \lambda_2\mu_1 - \lambda_2\mu_2]
-e_{\kappa}[-\lambda_1\mu_1 - 2\lambda_1\mu_2 - 2\lambda_2\mu_1 - \lambda_2\mu_2] \right) \;,
\end{eqnarray*}
where $e_{\kappa }[x]:=\exp [\frac{-2i\pi x}{3\kappa }]$ and where the vertices are labelled by shifted Dynkin labels
$\lambda = \{\lambda_1,\lambda_2\}$, $\mu = \{\mu_1,\mu_2\}$.
This $d_{{\cal A}_k}^2$ matrix $S$,  obtained as a --properly normalized and ordered -- quantum ``character table'' , defines the quantum analogue of a Fourier transform for the graphs ${\cal A}_k$.
The matrix $S$ is symmetric and such that $S^4 = 1$.
In the opposite direction, the well known Verlinde formula \cite{Verlinde} expresses fusion matrices $N_{\lambda}$
in terms of coefficients of $S$:
\begin{equation}
{\mathcal N}_{\lambda \mu}^{\nu} = \sum_{\beta \in \mathcal{A}_{k}}
\frac{S_{\lambda \beta} \, S_{\mu \beta } \, S_{\nu \beta}^{*}}{S_{0 \beta}} \; ,
\end{equation}
where $\lambda = 0 = (0,0)$ is the trivial representation.
In the present paper we prefer to obtain the $S$ matrix from  the combinatorial data provided by the graph.

\paragraph{The modular generator $T$}
The modular generator $T$ is diagonal in the basis defined by vertices.  Its eigenvalue associated with a vertex of shifted coordinates $\lambda  = \{\lambda_1,\lambda_2\}$ is equal to \cite{Kac-Peterson}:
\begin{equation}
T_{\lambda \lambda} = \exp \left[ 2 i \pi \left( \frac{[\lambda_1^2 + \lambda_1\lambda_2 + \lambda_2^2] - \kappa} {3\kappa }
\right) \right] \;.
\end{equation}
The square bracket in the numerator of the argument of $\exp$  can be simply read from the coordinates of the chosen vertex since it is the corresponding eigenvalue for the quadratic Casimir of  the Lie group $SU(3)$. We call ``modular exponent'' the whole numerator (\ie the difference between the Casimir and the generalized Coxeter value $\kappa$) taken modulo $3 \kappa$.
The $T$ operator is therefore essentially (up to a trivial geometric phase) obtained as the exponential of the quadratic Casimir:  the values for the shift ($ - \kappa $) and multiplicative constant ($3\kappa$) can indeed be fixed by imposing that the $SL(2,\ZZ)$ relation $(S T)^3 = S^2$ hold.

\paragraph{The $SL(2,\ZZ)$ representation defined by ${\cal A}_k$}
Matrices $S$ and $T$ provide therefore a representation of  the group $SL(2,\ZZ)$ for each alcove of $SU(3)$. Actually, one obtains moreover the identity $T^ {3\kappa} = 1$ so that
this representation factorizes through the finite group
$SL(2,\ZZ / 3 \kappa \ZZ)$.

\subsection{Symmetry and automorphism}

\paragraph{The $Z_3$ action}
Rotations of angle $0, 2 \pi/3$ or $4 \pi/3$ around 
the center of the equilateral triangle associated with
the graph ${\cal A}_k$ define a $Z_3$ action -- that we denote
by $z$ -- on the set of vertices and therefore an
endomorphism of the algebra (its cube is the identity). Its action on the irreps labelled by Dynkin labels
$(\lambda_1,\lambda_2)$ is given by:
\begin{equation}
z(\lambda _{1},\lambda _{2}) = (k-\lambda _{1}-\lambda _{2},\lambda _{1})\;.
\label{zauto}
\end{equation}

\paragraph{The Gannon automorphism $\rho$}

It is defined on the vertices, as \cite{Gannon-su3}
\begin{equation}
\rho = z^{k t}\;,
\label{gantwist}
\end{equation}
where $t$ is the triality and $k$ is the level of the graph.
We found the following result \cite{GilDahmaneHassan}:  if vertices $v_1$ and
$v_2$ are such that $v_2 = \rho [v_1] $, then
$T[v_1]=T[v_2]$.  The proof is given in \cite{GilDahmaneHassan}.


\section{General properties of the $SU(3)$ system of graphs}
This is a collection of graphs. As it will be discussed later,
each graph $G$ gives rise to a weak Hopf algebra (a quantum
groupo\"{\i}d) ${\cal B}G$, and each graph $G$ is also associated
with a given $\widehat{su}(3)$ modular invariant $Z$. At the
moment, we suppose that the collection of graphs (also called the
``Coxeter-Dynkin system of type $SU(3)$'') is given and we list
several of their properties. Several graphs (the orbifolds of the
A series) were obtained by Kostov \cite{Kostov} but the full list
of graphs for this system was obtained by Di Francesco and Zuber
\cite{DiFZuber,DiF}. Later, A. Ocneanu, at the Bariloche school
2000 \cite{Oc-Bariloche}, explained why one member of their
original list had to be removed.

\subsection{First properties}

\paragraph{Vertices and edges}
Vertices of $G$ are denoted $a,b,c,\ldots$. Edges are
oriented. In some cases there are multiple edges between
two vertices.

\paragraph{Spectral properties of the graph $G$}
A graph $G$ belonging to the $SU(3)$ system is
characterized by an adjacency matrix.
Its biggest eigenvalue is called $\beta = 1+2\cos(2\pi/\kappa)$. The
Coxeter number $\kappa$ is read from $\beta$. The level
is defined as $k=\kappa -3$. The set of eigenvalues of the graph $G$ is a
subset of the eigenvalues of the graph ${\cal A}_k$ with same level. They are of the form
$\beta(r_1,r_2)$ in Eq.(\ref{valprop}), with possible multiplicities.
The pairs of integers $(r_1,r_2)$ are called the exponents of the graph $G$.

\paragraph{The associated modular invariant}
$SU(3)$ graphs have been proposed as graphs associated to $\widehat{su}(3)$ modular invariant
partition functions. These partition functions $Z$ are sesquilinear forms on the characters
labelled by irreps of $\widehat{su}(3)_k$. The correspondance is such that diagonal
terms of $Z$ match the set of exponents for the corresponding graph $G$. The interpretation for
the off diagonal terms of Z was found by A. Ocneanu \cite{Oc-paths,Oc-opalg}.
We shall come back to this later.

\paragraph{Quantum dimensions and order of $G$}
One of the vertices of the graph $G$, denoted ${\bf 0}$, is called the unit vertex. It is defined from the eigenvector
corresponding to $\beta$ as the vertex associated to the smallest component\footnote{If the graph possesses some
(classical) symmetry, there can be several vertices associated to the smallest component. In those cases, we just
choose one of them.}. The components of the normalized eigenvector associated
with  $\beta$  (the dimension vector) define the quantum dimensions of the corresponding
vertices -- normalisation is obtained by setting to $1$ the quantum dimension of
the unit vertex\footnote{It plays indeed the role of a unit when the graph G has self-fusion (see later),
otherwise it is only a vertex whose quantum dimension is 1.}. 
When there is only one arrow leaving (and going to) the unit vertex\footnote{This is for instance not so for $\mathcal{D}_k^*$.} ${\bf 0}$, the quantum dimensions of its two neighbours (denoted as ${\bf 1}$ and ${\bf 1}^*$) are both equal to $\beta$. The sum of the squares of the quantum dimensions of vertices is called the order or the quantum mass of $G$, and
denoted $m(G)$.

\subsection{The two representation theories associated with the bialgebra ${\cal B}G$}
A quantum groupo\"{\i}d ${\cal B}G$ is associated with any graph
$G$ of the $SU(3)$ system. It is both semi-simple and co-semi-simple. We
present several basic properties here;  more details will be given
in Section $\bf{4}$.

\paragraph{The fusion algebra $A(G)$}
The algebra ${\cal B}G$ endowed with its associative product is a direct sum of
matrix algebras labelled by the index $\lambda$ (\ie by
vertices of the ${\cal A}_k$ graph with same level). Its representation theory (algebra
of characters) $A(G)$ is isomorphic to the fusion
algebra of ${\cal A}_k$.
Matrix representatives of the generators $\lambda$ of
${\cal A}_k$ have been already introduced: they correspond to the fusion
matrices $N_{\lambda}$.

\paragraph{The algebra of quantum symmetries $Oc(G)$}
 The dual algebra $\widehat{\cal B}G$ endowed with its associative
product is also a direct sum of matrix algebras labelled
by an index $x$.
Its representation theory (algebra  of characters) is
called the ``algebra of quantum symmetries'' of $G$ and denoted
$Oc(G)$. We call $d_O$ the dimension of $Oc(G)$.
It is  an algebra with a unit (denoted $0$) and,
for $SU(3)$ graphs, with -- in general but not always -- two algebraic generators (called
chiral left and chiral right generators and denoted as $1_L$ and $1_R$), together with their conjugates
$1_L^*$ and $1_R^*$. The Cayley graph of multiplication by the two
generators $1_L$ and $1_R$ (two types of lines) is called the Ocneanu
graph of G. The graph corresponding to the conjugated generators $1_L^*$ and $1_R^*$ is obtained from
the (oriented) Ocneanu graph by reversing the arrows.
$Oc(G)$ has also another conjugation, called the chiral
conjugation, that permutes the two algebraic generators $1_L$ and $1_R$.
Another way of displaying the Cayley graph is to draw only the graph of multiplication
by one chiral generator, say $1_L$, and to associate (for
example using dashed lines)  each  basis element with its
chiral conjugate. Multiplication of a vertex $x$ by the chiral generator $1_R$ is obtained as follows: we start with $x$, follow the dashed lines to find its chiral vertex $y$, then use the multiplication by $1_L$ and finally pull back using the dashed lines to obtain the result. Linear generators of $Oc(G)$ (\ie vertices of the Ocneanu graph)
that are identical with their chiral conjugates are called
self-dual. The two subalgebras generated by the chiral generators
are called chiral subalgebras.
The intersection of these two subalgebras is called the
ambichiral subalgebra, and its generators are the
ambichiral generators (they are self-dual).
$Oc(G)$, like $A(G) \simeq {\cal A}_k$, is not only an algebra
but an algebra that comes with a particular basis (the vertices of the Ocneanu graph),
for which structure constants are non negative integers. The multiplication
between vertices reads $x\,y = \sum_z O_{xy}^z\,z$, where $O_{xy}^z$, called
quantum symmetry coefficients, are non negative integers.
Matrix representatives of these linear generators $x$ of
$Oc(G)$ are called ``Ocneanu matrices''' and denoted $O_x$,
with elements $(O_x)_{yz} = O_{xy}^z$. They form an anti-representation of the Ocneanu
algebra:
\begin{equation}
O_x O_y = \sum_z O_{yx}^z O_z \;.
\end{equation}
If $Oc(G)$ is commutative - which is not always so -  then $O_{xy}^z=O_{yx}^z$ and the
Ocneanu matrices form a representation of the Ocneanu algebra: $O_x O_y =\sum_z O_{xy}^z O_z$.
The structure of  $Oc(G)$ is very much case dependent. One of the purpose of
this paper is actually to present the corresponding
results (for the $SU(3)$ system)  in a synthetic way.
In many cases $Oc(G)$ can be written as the direct sum of
a chiral subalgebra and one or several modules over this
subalgebra. Knowledge of the Ocneanu graph (\ie the action
of $1_L$ and $1_R$) may sometimes
be insufficient to encode the full structure (like for the $D_4$ case of the $SU(2)$ system).
Matrices $O_{1_L}$ and $O_{1_R}$
are the adjacency matrices of the Ocneanu graph. The two dimension vectors
(normalized eigenvectors associated with the largest
eigenvalue for each adjacency matrix) allow one to attribute -- unambiguously --  quantum dimensions to
all the linear generators of $Oc(G)$. Actualy, the two chiral generators have dimension $\beta$ and the whole list of quantum dimensions can be read directly from the Ocneanu graph by using the fact that this property is multiplicative
$qdim(x\,y) = qdim(x)\,qdim(y)$.
The sum of their squares is called the order or the quantum mass of $Oc(G)$, denoted $m(Oc(G))$: it is equal to the
order of $m(\mathcal{A}_k)$ of $\mathcal{A}_k = A(G)$. This property generalizes the usual group theory result.

\subsection{$G$ as a module over $A(G)={\cal A}_k$}
Call also $G$ the vector space spanned by the vertices of
a graph $G$. Call $r$ the number of vertices of the graph.
This vector space is a module for the action of the fusion algebra
associated with ${\cal A}_k$, where
$k$ is the level of $G$ (Coxeter number minus $3$).
The action is defined by the relation $ \lambda \, a = \sum_b F_{\lambda a}^b \, b$,
where $F_{\lambda a}^b$ are non negative integers called fused or annular coefficients.
In some cases, the same graph $G$ may also be a module over
some other graph of type $A$ with a different
Coxeter value, but we are not interested in this
phenomenon.

\paragraph{Annular matrices}
This action is encoded by a set of matrices $F_{\lambda}$ called  annular matrices or fused
(not fusion !) matrices, defined by $(F_{\lambda})_{ab} = F_{\lambda a}^b$.
From the module property $\lambda\, (\mu\,a) = (\lambda\,\mu)\,a$, the annular matrices satisfy:
\begin{equation}
F_{\lambda} \, F_{\mu} = \sum_{\nu} N_{\lambda\mu}^{\nu}\,F_{\nu}\;.
\label{FF}
\end{equation}
They form a representation of the fusion algebra (usually of different dimension since $r \not= d_{\mathcal{A}_k}$).
They are obtained by the same recurrence relation (\ref{recN}) as the fusion matrices but with
$F_{(0,0)} = \munite_{r \times r}$ and $F_{(1,0)} = Ad(G)$, where $Ad(G)$ is the adjacency matrix of
$G$. We obtain in this way $d_A$ matrices of size $r \times r$. As before $d_A$ is the number of vertices of the
associated ${\cal A}_k$ graph, the index $\lambda$ of $F_{\lambda}$ is a
Young tableau.

\paragraph{Essential paths (also called horizontal paths)}
Since annular matrices $F_{\lambda}$ have non negative integer
entries, one can associate a graph to every such matrix.
If the matrix element of $(F_{\lambda})_{ab} = p$, we
introduce $p$ oriented edges between vertices $a$
and $b$ of $G$. Such an edge is called an essential
path of type $\lambda$ from $a$ to $b$.
This graph will be called the horizontal graph of type $\lambda$.
Remember that the
$\lambda$ index is a Young tableau (a vertex of the
corresponding ${\cal A}_k$ diagram).
The graph associated with the generator $F_{(1,0)}$ is
the graph $G$ itself.

\paragraph{Essential matrices (or horizontal matrices)}
Essential matrices have the same information contents
as the annular matrices, however, they are
rectangular rather than square. They are defined as
follows
\begin{equation}
(E_a)_{\lambda b}  \doteq (F_{\lambda})_{ab}\;.
\label{essmatr}
\end{equation}
We have therefore one essential matrix $E_a$ for each
vertex $a$ of the graph $G$.
The integer $(E_a)_{\lambda b} $ gives the number of
horizontal  paths of type $\lambda$ from $a$ to $b$.
The property (\ref{FF}) can be written as follows using
essential matrices:
\begin{equation}
N_{\lambda}\,E_a = E_a\,F_{\lambda}\;.
\label{intertwin}
\end{equation}
In particular we have $N_{(1,0)}\,E_0 = E_0 \, F_{(1,0)}$. The essential matrix $E_0$ associated with the unit 0 of the graph $G$ intertwines the adjacency matrices of the graphs $G$ and ${\cal A}$:
it is also called the $({\cal A}_k,G)$ intertwiner.

\paragraph{Restriction-induction coefficients}
Non-zero entries of  the first line of $F_{\lambda}$ (ie relative to
the unit vertex of $G$) are called restriction coefficients.
They define a restriction
from ${\cal A}_k$ to $G$ (like irreps of a group versus
irreps of a subgroup). The branching rules are given by:
\begin{equation}
\lambda \hookrightarrow \sum_b (F_{\lambda})_{1b} \, b = \sum_b (E_0)_{\lambda b} \, b \;.
\end{equation}
The line indices corresponding to the non-zero
entries of the column $b$ of the matrix $E_0$ are
called induction coefficients associated with the
vertex $b$. They give the vertices $\lambda$ for which $b$ appears in their branching rules.
The line indices (Young tableaux)  corresponding to the non-zero
entries of the first column of the matrix $E_0$
are called degrees  of the family of would-be quantum invariants tensors by analogy with the situation that prevails for finite subgroups of Lie groups (for instance, when $G$ is the fusion graph by the fundamental representation of binary polyhedral groups,  these non-zero entries of the first column of $E_0$ reflect the existence of invariant symmetric tensors and therefore give the degrees of the Klein invariant polynomials for symmetry groups of Platonic bodies).

\subsection{$G$ as a module over $Oc(G)$}
The vector space $G$ is also a module for the action of the
algebra of quantum symmetries $Oc(G)$. Call $x$ the elements of
$Oc(G)$. The action is defined by the relation $ x \, a = \sum_b
S_{x a}^b \, b$, where $S_{x a}^b$ are non negative integers
called dual annular coefficients.

\paragraph{Dual annular matrices}
The action can be encoded in a set of matrices $S_x$ that we call
the dual annular matrices, defined by $(S_x)_{ab} = S_{x a}^b$.
From the module property $x\, (y\,a) = (x\,y)\,a$, the dual
annular matrices satisfy:
\begin{equation}
S_x \, S_y = \sum_{z} O_{yx}^{z}\,S_{z}\;.
\end{equation}
They satisfy the same relations as the Ocneanu matrices $O_x$ (they form an anti-representation
of the quantum symmetry algebra). We obtain in this way $d_O$ matrices of size $r \times
r$. As before $d_O$ is the number of vertices of the
associated Ocneanu graph.

\paragraph{Vertical paths}
Since dual annular matrices $S_x$ have non negative integer
entries, one can associate a graph to every such matrix.
 If the matrix element $(S_x)_{ab} = p$, we
introduce $p$ oriented edges between vertices $a$
and $b$ of $G$.  Such an edge is called a vertical
path of type $x$ from $a$ to $b$.
This graph will be called the vertical  graph of type $x$.
The vertical graphs associated with the two chiral generators of
$Oc(G)$ coincide with $G$ itself.

\paragraph{Vertical matrices}
Vertical matrices have the same information content as the dual
annular matrices, however, they are rectangular rather than
square. They are defined as follows:
\begin{equation}
(R_a)_{xb}  \doteq (S_x)_{ab} \;.
\end{equation}
We have therefore one vertical matrix $R_a$ for each
vertex $a$ of the graph $G$.
The integer $(R_a)_{xb}$ gives the number of vertical
paths of type $x$ from $a$ to $b$.

\subsection{Self-fusion}
${\cal A}_k$ diagrams have self-fusion (the fusion algebra). A
graph $G$ has self-fusion when the vector space spanned by its
vertices is not only a module over the corresponding $A(G)$ fusion
algebra but when it possesses an associative algebra structure
encoded by the graph itself (its adjacency matrix), with non negative
integral structure constants, compatible with the already known
$A(G)$ action. If $a,b,c,\ldots$ are vertices of a graph $G$ with
self-fusion, we have $a \, b =  \sum_c G_{ab}^c \, c$, where the
coefficients are non negative integers. The unit ${\bf 0}$ of the
graph is the identity for the multiplication. The multiplication
of some chosen vertex by the special vertex ${\bf 1}$ (resp. ${\bf
1^*}$) is given by the sum of vertices $a$ such that there is an
edge of $G$ from the chosen vertex to $a$ (resp. from $a$ to the
chosen vertex). The compatibility condition between self-fusion
and module structure reads $\lambda (a\,b) = (\lambda \,a)b$.

\paragraph{Conjugation}
Conjugation is defined for all self-fusion graphs. It
is compatible with the conjugation already defined for
$A$ graphs. We call $a^*$ the conjugate of $a$ in $G$. The
compatibility condition is understood as follows: all vertices of
${\cal A}_k$ appearing in the induction list associated with $a^*$
should be the conjugated vertices (taken in ${\cal A}_k$) of
those associated with $a$. When these two sets are equal, then $a^*=a$.
This provides a method for determining the conjugation of the $G$ vertices.
We have $(\lambda a)^*=\lambda^* a^*$, thus the annular coefficients should satisfy
$(F_{\lambda^*})_{a^*b^*} = (F_{\lambda})_{ab}$.

\paragraph{Triality}
Triality is also defined for all graphs with self-fusion. It
is compatible with the triality already defined for
$A$ graphs. This compatibility condition is
understood as follows: if the level of the graph $G$ is
$k$, then all the vertices of ${\cal A}_k$ appearing in the
induction list associated with a given vertex of $G$
should have the same triality. This provides a method for
determining the triality of the $G$ vertices.

\paragraph{Graph matrices}
The fusion of $G$ vertices can be encoded in a set of matrices
$G_a$ with non negative integer coefficients $(G_a)_{bc} = G_{ab}^c$, called
graph matrices. We have $G_0 = F_{(0,0)}$,
$G_{1} = F_{(1,0)}$ and $G_{1^*} = F_{(0,1)}$.
The compatibility condition for graphs with
self-fusion (cf supra) reads $G_a\, F_{\lambda} = F_{\lambda}\,G_a$.
In particular, using essential matrices $E_a$ defined in Eq.(\ref{essmatr}) one can get
$E_a = E_0\,G_a$.

\paragraph{Remark}
Some of the graphs belonging to a Coxeter-Dynkin
system have self-fusion, others don't. For example,
in the $SU(2)$ system, the diagrams $A_n$,
$D_{even}$, $E_6$ and $E_8$ have self-fusion, this is
not the case for $D_{odd}$ and $E_7$. In the $SU(3)$ system,
diagrams ${\cal A}_k$, ${\cal D}_{3n}$, ${\cal E}_5$, ${\cal E}_9$ and
${\cal E}_{21}$ have self-fusion. The others don't.

\paragraph{Flatness}
We believe that self-fusion is equivalent to
flatness, as defined for instance in \cite{Oc-string,Oc-qsym} or
\cite{sunder}. The two notions look a priori very different
but it seems that all known graphs with self-fusion
are also flat (and reciprocally).  We are not aware
of any formal proof relating the two concepts.

\subsection{Coxeter-Dynkin systems of graphs, self-connections and
Kuperberg spiders} A graph that is a member of a Coxeter-Dynkin
system gives  rise to a particular kind of quantum groupo\"{\i}d.
Such a graph is associated with some modular invariant, but
sometimes more than one graph can be associated with the same
invariant. Moreover, a member of a Coxeter-Dynkin system has also
to  be compatible, in a sense that should be precised, with a
given Lie group (here $SU(3)$). Being a module over the graph
algebra of a Weyl alcove at some level is a necessary but not
sufficient condition. A condition, using the notion of
self-connections on graphs, was given by A. Ocneanu in Bariloche
(2000) \cite{Oc-Bariloche} and this lead him to discard one of the
graphs of the original Di Francesco - Zuber list. We believe that
the appropriate algebraic concept can be phrased in terms of
Kuperberg spiders \cite{KuperbergSpiders} but we have no rigorous
proof that the two concepts are the same.


\section{The quantum groupo\"{\i}d associated to a pair ($G_1,G_2$)}
If $G_{1}$ has self-fusion and if $G_{2}$ is a module over
$G_{1}$, one can associate a bialgebra ${\cal B}(G_{1},G_{2})$ to
this pair of graphs \cite{Oc-paths}. This bialgebra is a
particular type of weak Hopf algebra (or quantum groupo\"{\i}d)
(see for instance
\cite{Oc-Guadeloupe,Bohm-Szc,Bohm-WHA,Nill,Vainerman}). We call it
the ``Ocneanu quantum groupo\"{\i}d'' associated with the chosen
pair. In particular if $G_{2}=G$ and $G_{1}=\mathcal{A}_{k}$, with
$k$ the level of $G$, we just denote ${\cal B}G\doteq {\cal
B}(\mathcal{A}_{k},G)$, or simply ${\cal B}$ if the choice of $G$
is clear from the context. In what follows we consider mostly
bialgebras of that type.

\subsection{The vector spaces $\mathcal B$ and $\widehat{\mathcal B}$}
\paragraph{Admissible triangles}
To every essential (i.e. horizontal) path of type $\lambda$
between $a$ and $b$ one associates a triangle with one
horizontal edge  labelled by $\lambda$ and two edges labelled
by $a$ and $b$. Such triangles (with
$1$ line of type $A$ and $2$ lines of type $G$) are
called admissible triangles. By duality, they can also be
drawn as $(GGA)$ vertices.
The vector space spanned by such triangles is called
$EssPath(G)$ or $Hpaths(G)$, it is graded by $\lambda$:
$Hpaths(G) = \sum_{\lambda}  Hpaths_{\lambda}(G)$.

To every vertical path of type $x$ between $a$ and $b$
one associates a triangle with one vertical edge
labelled by $x$ and two edges labelled by $a$ and $b$.
Such triangles (with $1$ line of type $Oc$ and $2$ lines of
type $G$) are also called admissible triangles. By duality,
they can also be drawn as $(GGO)$ vertices.
The vector space spanned by such triangles is called
$Vpaths(G)$, it is graded by $x$ : $Vpaths(G) = \sum_x
Vpaths_x(G)$.

\begin{figure}[h]
\setlength{\unitlength}{0.85mm}
\par
\begin{center}
\begin{picture}(120,20)

\bezier{30}(25,0)(22.5,-2)(17.5,-6)\bezier{30}(25,0)(27.5,-2)(32.5,-6)
\bezier{30}(25,15)(22.5,17)(17.5,21)\bezier{30}(25,15)(27.5,17)(32.5,21)
\put(15,-6){\makebox(0,0){$c$}} \put(35,-6){\makebox(0,0){$d$}}
\put(15,21){\makebox(0,0){$a$}} \put(35,21){\makebox(0,0){$b$}}
\put(27,7){\makebox(0,0){$\lambda$}}

\put(24.5,0.5){$\wr$}\put(24.5,3.5){$\wr$}\put(24.5,6.5){$\wr$}\put(24.5,9.5){$\wr$}
\put(24.5,12.5){$\wr$}

\put(52,7){$\equiv$}\put(70,7.5){\line(1,0){15}}
\put(70,7.5){\line(1,2){7.5}}\put(70,7.5){\line(1,-2){7.5}}
\put(85,7.5){\line(-1,2){7.5}}\put(85,7.5){\line(-1,-2){7.5}}

\put(71,15){\makebox(0,0){$a$}} \put(84,15){\makebox(0,0){$b$}}
\put(72,-3){\makebox(0,0){$c$}} \put(83,-3){\makebox(0,0){$d$}}
\put(77,10){\makebox(0,0){$\lambda$}}

\put(70,7.5){\circle*{4}} \put(85,7.5){\circle*{4}}
\put(77.5,22.5){\circle{4}} \put(77.5,-7.5){\circle{4}}

\end{picture}
\end{center}
\caption{A double triangle of type GGAGG of ${\cal B}$.}
\label{Fig:DoubleTriangleA}
\end{figure}
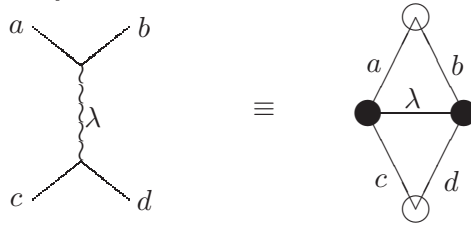

\paragraph{Double triangles}
We call ${\cal B}$ the graded vector space $\sum_{\lambda}  Hpaths_{\lambda}(G)
\otimes Hpaths_{\lambda}(G)$. It is spanned by double triangles $GGAGG$
(two triangles of type $(GGA)$ sharing a common edge  of
type $A$). By duality they can also be drawn as diffusion
diagrams (like in Figure \ref{Fig:DoubleTriangleA}).

\begin{figure}[h]
\setlength{\unitlength}{0.85mm}
\par
\begin{center}
\begin{picture}(120,25)

\put(15,10){\line(1,0){20}}\put(15,10.25){\line(1,0){20}}\put(15,9.75){\line(1,0){20}}
\bezier{30}(15,10)(12.5,8)(7.5,4)\bezier{30}(15,10)(12.5,12)(7.5,16)
\bezier{30}(35,10)(37.5,8)(42.5,4)\bezier{30}(35,10)(37.5,12)(42.5,16)
\put(7,19){\makebox(0,0){$a$}}
\put(7,1){\makebox(0,0){$c$}} \put(42,19){\makebox(0,0){$b$}}
\put(42,1){\makebox(0,0){$d$}} \put(25,13){\makebox(0,0){$x$}}

\put(60,10){$\equiv$}
\put(87.5,25){\line(0,-1){30}}\put(80,10){\line(1,2){7.5}}\put(80,10){\line(1,-2){7.5}}
\put(95,10){\line(-1,2){7.5}}\put(95,10){\line(-1,-2){7.5}}

\put(81,17.5){\makebox(0,0){$a$}}
\put(94,17.5){\makebox(0,0){$b$}} \put(82,0){\makebox(0,0){$c$}}
\put(93,0){\makebox(0,0){$d$}} \put(89,10){\makebox(0,0){$x$}}

\put(80,10){\circle*{4}} \put(95,10){\circle*{4}}
\put(87.5,25){\circle{4}} \put(87.5,-5){\circle{4}}

\end{picture}
\end{center}
\caption{A double triangle of type GGOGG of $\widehat{\cal B}$.}
\label{Fig:DoubleTriangleO}
\end{figure}
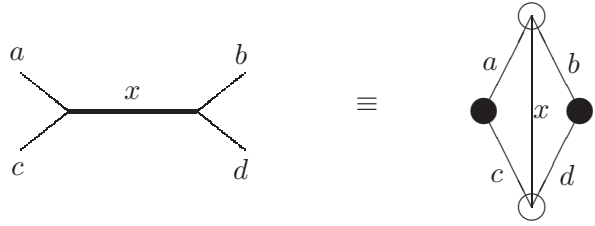

We call $\widehat {\cal B}$ the graded vector space $\sum_x
Vpaths_x(G) \otimes Vpaths_x(G)$. It is spanned by double
triangles $GGOGG$ (two triangles of type $(GGO)$ sharing a common
edge  of type $O$). By duality they can also be drawn as
diffusion diagrams (like in Figure \ref{Fig:DoubleTriangleO}).

\subsection{The multiplications}

\paragraph{The multiplication $\circ$ on the vector space ${\cal B}$}
This algebra structure on ${\cal B}$ is obtained by choosing
the set of double triangles of type $(GGAGG)$ as a basis
of matrix units $e_{IJ}$ for an associative product that we
call $\circ$, and such that multi-indices are like $\{I,J\} = \{(\lambda, a,b),(\lambda,c,d)\}$, i.e. 
with same $\lambda$.

\paragraph{The multiplication $\hat \circ$ on the dual vector space
$\hat {\cal B}$}
This algebra structure on $\hat {\cal B}$ is obtained by
chosing the set of double triangles of type $(GGOGG)$
as a basis of matrix units $\epsilon^{AB}$
for an associative product that we call $\hat \circ$, and such that
multi-indices are like  $\{A,B\} = \{(x, a,b),(x,c,d)\}$, i.e. with same $x$. 

\paragraph{Comultiplications and compatibility : Ocneanu cells}
Since we have a product $\circ$ in ${\cal B}$ we have a
coproduct $\hat \Delta$ in $\widehat {\cal B}$.
Since we have a product $\hat \circ$ in $\widehat {\cal B}$
we have a coproduct $\Delta$ in ${\cal B}$.  In
order to have a bialgebra structure, we need a
compatibility condition for the coproducts
(homomorphism property).  In order to ensure this,
it is not possible to assume that the two bases of
double triangles that we have used in ${\cal B}$ and in
$\widehat {\cal B}$ are dual bases. At the contrary, the
fact that there exists a non trivial pairing (between
these two bases) such that the compatibility
conditions holds is the main non trivial part of the
claim that ${\cal B}$ is actually a bialgebra. This non
trivial pairing is determined by the family of
Ocneanu cells or inverse cells $ < \epsilon^{AB}, e_{IJ} > $,
labelled with tetrahedra $a,b,\lambda, d,c, x$ (in some cases there is more than one
path -- horizontal or vertical --with fixed $\lambda$ or $x$ and given endpoints, so that  cells
may depend of other indices).
Explicit determination of these numerical coefficients is not
studied in the present paper.

For an arbitrary graph $G$,
there are actually several (five) sets of such
coefficients generalizing the Racah-Wigner $6j$
symbols; they obey orthogonality relations and
several types (five) of mixed pentagonal relations.
Their proper definition involves non-trivial
normalization choices.

\paragraph{Scalar product and convolution product}
Making a particular choice for a scalar product in
${\cal B}$, it is possible to trade the associative product
$\hat \circ$, defined on the dual vector space
$\widehat {\cal B}$ against an associative product $\ast$
(convolution product) in the vector space ${\cal B}$.
The situation is self-dual so that we can also find a scalar product in $\widehat {\cal B}$ in order to  trade the associative product
$\circ$ defined on ${\cal B}$ against an associative product $\hat \ast$ in the dual vector space
$\widehat {\cal B}$.

\subsection{Properties of $B$}
It is a finite dimensional semi-simple algebra and
co-semi-simple coalgebra (equivalently, its dual
$\widehat {\cal B}$ is also a finite dimensional semi-simple
algebra and co-semi-simple coalgebra).

\paragraph{Quadratic sum rules}
We call $d_{\lambda}=\dim(HPath_{\lambda})$ the dimensions of the
blocks labelled by $\lambda$,  associated with the first algebra
structure,  and $d_x = \dim(VPath_x)$ the dimensions of those
labelled by $x$, associated with the other algebra structure.
Since the underlying vector space is the same, and since both
algebra structures are semi-simple, we can calculate the dimension
$d_{\cal B}$ of ${\cal B}$ in two possible ways and check the
identity:
\begin{equation}
d_{\cal B} = \sum_{\lambda} d_{\lambda}^2 = \sum_x d_x^2 \;.
\end{equation}
The dimensions $d_{\lambda}$ and $d_x$ can be calculated from the
annular and dual annular matrices: $d_{\lambda} = \sum_{a,b}
(F_{\lambda})_{ab}$, $d_x = \sum_{a,b} (S_x)_{ab}$.

\paragraph{Linear sum rules}
Call $d_H = \sum_{\lambda}  d_{\lambda}$ and $d_V = \sum_x
d_x$. It happens that, in many cases, the
relation $d_H = d_V$ holds, and when it does not, one
knows how to correct it. Existence of this linear sum
rule (first observed in \cite{Pet_Zub-Occells}) is an
observational fact. Its origin is not understood.

\paragraph{${\cal B}$ is not a Hopf algebra but a weak Hopf algebra (a
quantum groupo\"{\i}d)} The main difference with the quantum group
case is that the coproduct of the unit is not equal to the tensor
square of the unit.  What replaces it can be written  $\sum
\munite_{(1)} \otimes \munite_{(2)}$. The terms appearing in this
sum also show up in the axioms defining weak Hopf algebras (see
for instance \cite{Bohm-Szc}). In particular the appropriate
tensor product for the category of representations is not
$\otimes$ but $\otimes \circ \Delta \munite \,$.

\paragraph{Available references}
The fact that a quantum groupo\"{\i}d  is associated with every
member of a Coxeter-Dynkin system is not phrased as such in
\cite{Oc-paths} but the two multiplicative structures are described
there in quite general terms\footnote{This description is clearly
related to the concept of (Ocneanu) paragroups introduced a long
time before the notion of quantum groupo\"{\i}d.}. The
correspondance between ADE  graphs and particular weak Hopf
algebras  is also strongly  suggested in \cite{Pet_Zub-Occells}.
Nowadays the fact that any member of a Coxeter-Dynkin system is
associated with a quantum groupo\"{\i}d (as defined by
\cite{Bohm-Szc}) belongs to the folklore (see \cite{ostrik,Cal_Etingof} for the description 
of this situation in the language of fusion categories and module categories). They are actually
quantum groupo\"{\i}ds of a very particular kind (so they should
better be called ``Ocneanu quantum groupo\"{\i}ds''). In the
case of the $SU(2)$ system, elementary proofs, based on axiomatic
properties of Ocneanu cells, are now available in published form
\cite{Coq_Trinchero}; several explicit examples have also been
worked out (for instance in \cite{Coq-mexico} or
\cite{Gil-thesis}). In the case of the $SU(3)$ system, general
proofs are not available. Our attitude in this paper is however to
take the above property for granted.


\section{The double fusion algebra and the modular splitting}

\subsection{Bimodule properties}

\paragraph{Toric matrices and double annular matrices}
The Ocneanu quantum groupo\"{\i}ds  $\mathcal{B}G$ are of a very
special kind. In particular, we have the following property
involving simultaneously the two representation theories
associated with the bialgebra $\mathcal{B}G$ -- the fusion algebra
$A(G)$ and the quantum symmetries algebra $Oc(G)$ : $Oc(G)$ is an
$A(G)$ bimodule, \ie an $A(G) - A(G)$ module. This comes from the
fact that in all cases, $Oc(G)$ can be written as the tensor
square (maybe twisted or quotiented) of some graph algebra on
which $A(G)$ acts. We write this action $\lambda \, x \, \mu =
\sum_y (V_{\lambda\mu})_{xy}\, y$. The $V_{\lambda \mu}$ are $d_O
\times d_O$ matrices with non negative integer coefficients,
called double annular matrices. The same information can be
encoded in $d_{{\cal A}_k} \times d_{{\cal A}_k}$ matrices
$W_{xy}$ called toric matrices, with non negative integer
coefficients defined by $(W_{xy})_{\lambda \mu} \doteq (V_{\lambda
\mu})_{xy}$.

\paragraph{Double fusion equation}
The bimodule associativity property $(\lambda \lambda') x (\mu \mu') =
\lambda (\lambda' x \mu) \mu'$ leads to the following equation, called the double fusion equation:
\begin{equation}
V_{\lambda\mu} \, V_{\lambda'\mu'} = \sum_{\lambda''\mu''} \, N_{\lambda\lambda'}^{\lambda''} \,
N_{\mu\mu'}^{\mu''}\, V_{\lambda''\mu''}\;.
\label{DFE}
\end{equation}
This equation taken at $\mu=\mu'=0$, at $\lambda=\lambda'=0$ and at $\lambda' = \mu=0$ leads to:
\begin{eqnarray}
V_{\lambda 0} \, V_{\lambda' 0} &=& \sum_{\lambda''} N_{\lambda\lambda'}^{\lambda''} \, V_{\lambda'' 0}
\\
V_{0 \mu} \, V_{0 \mu'} &=& \sum_{\mu''} N_{\mu\mu'}^{\mu''} \, V_{0\mu''} \\
V_{\lambda\mu'} &=& V_{\lambda 0} \, V_{0 \mu'}  = V_{0\mu'} \, V_{\lambda 0} \;.
\end{eqnarray}
Each set of matrices $V_{\lambda 0}$ or $V_{0 \mu}$ gives therefore a representation of dimension $d_O \times d_O$ of the fusion algebra and $V_{0 0}$ is the identity matrix. They can be determined by the
same recurrence relation as the fusion matrices $N_{\lambda}$, once the fundamental generators
 $V_{(1,0),(0,0)}$ and $V_{(0,0),(1,0)}$ are known.

\paragraph{Other properties of $V_{\lambda\mu}$ matrices}
The action is central. Writing $\lambda (x \, y) \mu = x(\lambda  y \mu) = (\lambda x  \mu)y $ leads to:
\begin{equation}
O_x \, V_{\lambda \mu} = V_{\lambda \mu} \, O_x = \sum_z (V_{\lambda \mu})_{xz} \, O_z \;.
\label{OVrel}
\end{equation}

\paragraph{The Ocneanu graph} With the set of relations satisfied by $V_{\lambda\mu}$ matrices and with the help of the known recurrence relations of irreps of $SU(3)$, all the coefficients $(V_{\lambda\mu})_{xy}$ can be simply determined from the fundamental matrices $V_{(1,0),(0,0)}$ and $V_{(0,0),(1,0)}$. These matrices are the adjacency matrices of the Ocneanu graph:
\begin{equation}
V_{(1,0),(0,0)} = O_{1_L} \qquad \qquad V_{(0,0),(1,0)} = O_{1_R} \;.
\end{equation}
The Ocneanu graph determines (and is determined by) these two matrices.

\paragraph{Generalized partition functions}
In the boundary  conformal field theory
associated to the given  graph, the partition function on a torus with defect lines labelled by $x$ and $y$  is given by $Z_{xy} = \overline \chi \, W_{xy} \, \chi$ where $\chi$ is the vector of characters
of affine $su(3)$ \cite{Pet_Zub-gener}.

\paragraph{The modular matrix $M$}
In particular, when there are no defect lines ($x=y=0$), we recover the modular invariant partition function
$Z = \overline \chi \, M \, \chi$, since the modular invariant matrix $M=W_{00}$ commutes with the modular generators $S$ and $T$ in the  representation of $SL(2,\ZZ)$ associated with the Weyl alcove at this level.
In  contrast, the $V_{00}$ matrix is the identity matrix.

\paragraph{The double intertwining relation}
From the fact that a graph $G$ with level $k$ is an ${\mathcal
A}_k$ module we deduced the intertwining relation given in Eq.(\ref{intertwin}), written in terms
of essential matrices $E_a$ attached to each vertex of the graph
$G$. By analogy, let us introduce here the ``essential tensor'' $K_x$, with
components $(K_x)_{\lambda \mu y} = (V_{\lambda \mu})_{xy}$,
associated to each vertex $x$ of $Oc(G)$. It can be written as a
rectangular matrix of size $d_A^2 \times d_O$ (call it double
essential matrix). From the fact that $Oc(G)$ is an $A(G)$
bimodule, the double fusion equation (\ref{DFE}) can be written
using $K_x$, leading to the following double intertwining
relation:
\begin{equation}
\tau \circ (N_\lambda \otimes N_\mu) K_x = K_x \, V_{\lambda \mu} \;,
\label{DIR}
\end{equation}
where $\tau$ gives a flip on tensor components: $\tau \circ
(T_{(\lambda' \lambda'')\, (\mu' \mu'')}) = T_{(\lambda' \mu')\,
(\lambda''\mu'')})$.

\paragraph{Other useful formulae}

We already recalled the graph interpretation for the  diagonal entries of $M$ in terms of exponents of the graph.
More generally we have the following result \cite{Oc-paths,Oc-opalg}.
The number of vertices $d_O$ of the Ocneanu graph (also called ``number of irreducible quantum symmetries'') is equal
to the
sum of square of entries of the modular matrix. Moreover, the algebra of quantum symmetries is isomorphic to a direct
sum of finite dimensional matrix algebras of the form $\bigoplus_{m,n} Mat_{M_{mn}}(C)$ where $M_{mn}$ are the entries of
the modular matrix. In other words these entries  give the dimensions of the irreducible representations of this algebra.

Another interpretation for these numerical entries can be given in terms of  higher quantum Klein invariants (cf supra).

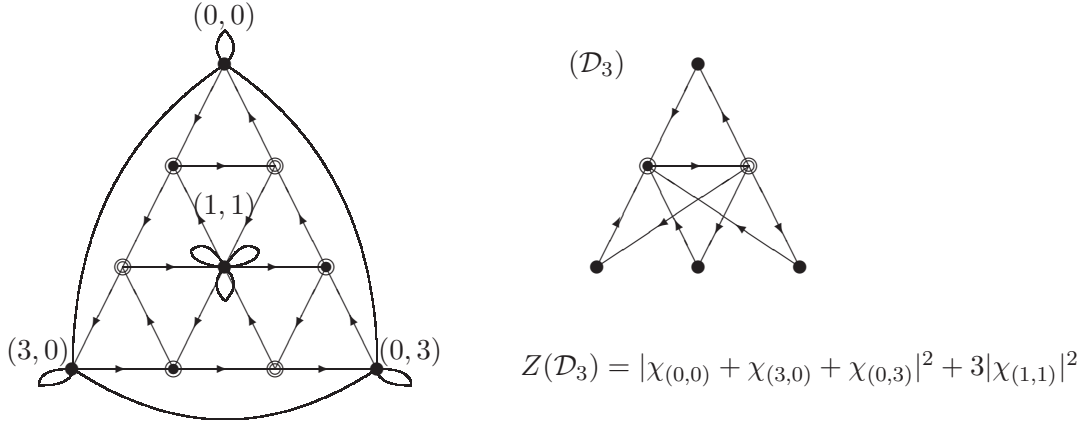
\begin{figure}[h]
\setlength{\unitlength}{0.9mm}
\begin{center}
\begin{picture}(150,70)


\put(7.5,5){\begin{picture}(15,15) \put(0,0){\line(1,0){15}}
\put(0,0){\line(1,2){7.5}} \put(15,0){\line(-1,2){7.5}}
\put(7.5,0){\vector(1,0){0.5}} \put(11.5,7){\vector(-1,2){0,5}}
\put(3.5,7){\vector(-1,-2){0,5}}
\end{picture}}

\put(22.5,5){\begin{picture}(15,15) \put(0,0){\line(1,0){15}}
\put(0,0){\line(1,2){7.5}} \put(15,0){\line(-1,2){7.5}}
\put(7.5,0){\vector(1,0){0.5}} \put(11.5,7){\vector(-1,2){0,5}}
\put(3.5,7){\vector(-1,-2){0,5}}
\end{picture}}

\put(37.5,5){\begin{picture}(15,15) \put(0,0){\line(1,0){15}}
\put(0,0){\line(1,2){7.5}} \put(15,0){\line(-1,2){7.5}}
\put(7.5,0){\vector(1,0){0.5}} \put(11.5,7){\vector(-1,2){0,5}}
\put(3.5,7){\vector(-1,-2){0,5}}
\end{picture}}


\put(15,20){\begin{picture}(15,15) \put(0,0){\line(1,0){15}}
\put(0,0){\line(1,2){7.5}} \put(15,0){\line(-1,2){7.5}}
\put(7.5,0){\vector(1,0){0.5}} \put(11.5,7){\vector(-1,2){0,5}}
\put(3.5,7){\vector(-1,-2){0,5}}
\end{picture}}

\put(30,20){\begin{picture}(15,15) \put(0,0){\line(1,0){15}}
\put(0,0){\line(1,2){7.5}} \put(15,0){\line(-1,2){7.5}}
\put(7.5,0){\vector(1,0){0.5}} \put(11.5,7){\vector(-1,2){0,5}}
\put(3.5,7){\vector(-1,-2){0,5}}
\end{picture}}


\put(22.5,35){\begin{picture}(15,15) \put(0,0){\line(1,0){15}}
\put(0,0){\line(1,2){7.5}} \put(15,0){\line(-1,2){7.5}}
\put(7.5,0){\vector(1,0){0.5}} \put(11.5,7){\vector(-1,2){0,5}}
\put(3.5,7){\vector(-1,-2){0,5}}
\end{picture}}

\qbezier[100](30,50)(27.5,52.5)(30,55)\qbezier[100](30,50)(32.5,52.5)(30,55)
\qbezier[1000](30,50)(7.5,35)(7.5,5)\qbezier[1200](7.5,5)(30,-10)(52.5,5)
\qbezier[1000](52.5,5)(52.5,35)(30,50)\qbezier[120](7.5,5)(2.5,5)(2.5,2.5)
\qbezier[120](7.5,5)(7.5,2.5)(2.5,2.5)\qbezier[120](52.5,5)(57.5,5)(57.5,2.5)
\qbezier[120](52.5,5)(52.5,2.5)(57.5,2.5)
\qbezier[100](30,20)(32.5,24.5)(35,22.5)\qbezier[100](30,20)(35,20.5)(35,22.5)
\qbezier[100](30,20)(27.5,24.5)(25,22.5)\qbezier[100](30,20)(25,20.5)(25,22.5)
\qbezier[100](30,20)(32.5,17.5)(30,15)\qbezier[100](30,20)(27.5,17.5)(30,15)


\put(30,50){\circle*{2}} \put(30,20){\circle*{2}}
\put(7.5,5){\circle*{2}} \put(52.5,5){\circle*{2}}

\put(22.5,35){\circle*{1.5}} \put(22.5,35){\circle{2.5}}
\put(45,20){\circle*{1.5}} \put(45,20){\circle{2.5}}
\put(22.5,5){\circle*{1.5}} \put(22.5,5){\circle{2.5}}

\put(37.5,35){\circle{2.5}} \put(37.5,35){\circle{1.5}}
\put(15,20){\circle{2.5}} \put(15,20){\circle{1.5}}
\put(37.5,5){\circle{2,5}} \put(37.5,5){\circle{1.5}}

\put(30,57){\makebox(0,0){$(0,0)$}}
\put(30,29){\makebox(0,0){$(1,1)$}}
\put(2.5,7.5){\makebox(0,0){$(3,0)$}}
\put(57.5,7.5){\makebox(0,0){$(0,3)$}}


\put(85,20){\begin{picture}(15,15) \put(0,0){\line(1,2){7.5}}
\put(0,0){\line(3,2){22.5}} \put(3.5,7.5){\vector(1,2){0}}
\put(9,6){\vector(-3,-2){0}}
\end{picture}}

\put(100,20){\begin{picture}(15,15) \put(15,0){\line(-1,2){7.5}}
\put(15,0){\line(-3,2){22.5}} \put(12.5,5){\vector(1,-2){0}}
\put(6,6){\vector(-3,2){0,5}}
\end{picture}}

\put(92.5,35){\begin{picture}(15,15) \put(0,0){\line(1,0){15}}
\put(0,0){\line(1,2){7.5}} \put(15,0){\line(-1,2){7.5}}
\put(0,0){\line(1,-2){7.5}}\put(15,0){\line(-1,-2){7.5}}
\put(7.5,0){\vector(1,0){0.5}} \put(11.5,7){\vector(-1,2){0,5}}
\put(4,8){\vector(-1,-2){0,5}}\put(10.5,-9){\vector(-1,-2){0,5}}
\put(4.5,-9){\vector(-1,2){0,5}}
\end{picture}}

\put(115,20){\circle*{2}} \put(100,50){\circle*{2}}
\put(100,20){\circle*{2}} \put(85,20){\circle*{2}}

\put(92.5,35){\circle*{1.5}} \put(92.5,35){\circle{2.5}}
\put(107.5,35){\circle{1.5}} \put(107.5,35){\circle{2.5}}

\put(85,50){\makebox(0,0){$(\mathcal{D}_{3})$}}

\put(115,5){\makebox(0,0){$Z(\mathcal{D}_{3})=|\chi_{(0,0)}+\chi_{(3,0)}+\chi_{(0,3)}|^2+3|\chi_{(1,1)}|^2$}}

\end{picture}
\end{center}
\caption{The modular diagram and the modular invariant associated
to the $\mathcal{D}_{3}$ graph}
\label{fig:modulardiagram}
\end{figure}

The above result was stated, by A. Ocneanu, for the $SU(2)$ system. It can also be checked explicitly 
for all members of the $SU(3)$ system. In the framework
of the theory of sectors, such a decomposition has been proved
in theorem 6.8 of \cite{bek} in a completely general setting (theorem 5.3 of the same paper shows that
it is equivalent to Ocneanu's graphical method). A nice graphical way to encode the
modular matrix $M$ associated with a graph $G$ is provided by the
``modular diagram'': it is a picture of the Weyl chamber at the
given level, with arcs connecting the vertices associated with
non-zero entries   $M_{mn}$. The degrees of quantum invariant tensors can also be read
from this diagram: they correspond to those vertices that belong to the arc going though the origin (0,0).
For instance figure
\ref{fig:modulardiagram} shows these results for the
$\mathcal{D}_{3}$ case.

The first part of the previous theorem can be written  $d_O = Tr (M \, M^{\dag})$.
When the modular splitting technique (see the next section) is used to determine explicitly the $W_{xy}$ and the algebra
$Oc(G)$ itself, the above result\footnote{Here -- and in the whole paper -- we have in mind the simply laced cases (the $ADE$ diagrams) or their generalizations.} provides a numerical check.

\subsection{Modular splitting}
The double fusion equation (\ref{DFE}) at $x=y=0$ leads to the following equation, written in terms of $W$ matrices, called the modular splitting equation :
\begin{equation}
\sum_z (W_{0z})_{\lambda\mu} \, (W_{z0})_{\lambda'\mu'} = \sum_{\lambda''\mu''} (N_{\lambda})_{\lambda'\lambda''} \,
(N_{\mu})_{\mu'\mu''} \, M_{\lambda''\mu''} \;.
\label{MSE}
\end{equation}
The double fusion equation (\ref{DFE}) at $y=0$ leads to the following equation, written in terms of $W$ matrices, called the generalized modular splitting equation:
\begin{equation}
\sum_z (W_{xz})_{\lambda\mu} \, (W_{z0})_{\lambda'\mu'} = \sum_{\lambda''\mu''} (N_{\lambda})_{\lambda'\lambda''} \,
(N_{\mu})_{\mu'\mu''} \, (W_{x0})_{\lambda''\mu''} \;.
\label{GMSE}
\end{equation}

\paragraph{Modular splitting technique $I$ : from the modular matrix $M$ to the
toric matrices $W_{x0}$}

In many cases, the graph $G$ itself is not known (see
comments in the last section) and the only knowledge that
we have is the modular matrix $M$. It is possible to use
the modular splitting equation to determine the toric
matrices. This was certainly the road followed by A. Ocneanu but a
general method of resolution  was first described in
\cite{EstebanRobert},  many more details and examples
can be found in \cite{EstebanGil}.

One starts from the modular splitting equation (\ref{MSE}). The
fusion matrices $N_{\lambda}$ and the modular matrix $M$ are
known. The right hand side of (\ref{MSE}) is thus known: it can be
seen as a matrix, called $K$ (the ``fused modular matrix''), of
size $d_A^2 \times d_A^2$.  Toric matrices that appear on the left
hand side are integer entries matrices $d_A \times d_A$ to be
determined. The number of {\sl distinct} toric matrices with one
twist is equal to the rank of $K$. In simple cases, the number
$d_O = Tr (M M^\dag)$ of Ocneanu generators $O_x$ is precisely
equal to the rank of $K$.  In more complicated cases the rank of
$K$ is strictly smaller (which means that several toric matrices
associated with distinct generators $O_x$ may coincide). The
explicit method leading to the determination of toric matrices
(\ie the technique used to solve  the modular splitting equation)
is not explained in the present paper.  It is described (for a
particular example) in one section of  \cite{EstebanRobert}. A
detailed study of this method together with several $SU(3)$
examples will be given in \cite{EstebanGil}.

\paragraph{Modular splitting technique $II$ : from the toric matrices
$W_{x0}$ to the Ocneanu generators $O_x$} Once we have determined
the toric matrices with one twist $W_{x0}$, we have to determine
the toric matrices $W_{xy}$. The right hand side of the
generalized modular splitting equation(\ref{GMSE}) is known. Toric
matrices $W_{xy}$ appearing on the left hand side can then be
calculated. This is equivalent to solve the double intertwining
relation (\ref{DIR}) in the particular case $x=0$ (this is a set
of linear equations that involves only the already determined
toric matrices with only one twist). This leads therefore  to the
determination of the  double annular  matrices and in particular
of the two chiral generators $O_{1_L}$ and $O_{1_R}$. The other Ocneanu
generators $O_x$ can be determined solving Eq. (\ref{OVrel}).

\paragraph{Remark}
Once the algebra (or graph) of quantum symmetries $Oc(G)$ has been obtained, we can 
determine the generalized Dynkin diagram $G$ as a module graph on $Oc(G)$. Sometimes 
there is not unicity of the result and two different graphs may be associated with
the same initial modular invariant. See also our comments in the last section.

\paragraph{Relative modular splitting formula and relative double fusion
algebra}
Often, the algebra $Oc(G)$ is not only a bimodule over
$A(G)$ but also a bimodule
over the graph algebra of $H$ where $H$ is a graph with
self-fusion on which $A(G)$ acts.  In the cases where $G$
admits self-fusion, it is often so that  $H$ is $G$
itself.
In those cases we have a relative modular splitting
formula: fusion matrices are still the same but the relative
modular matrix $M^{rel}$ is written in terms of the $G$ graph (so
it is of size $d_G^2$ rather than $d_A^2$);  $M = E_0 \,
M^{rel} \, E_0^T$, where $E_0$ is the first essential matrix
(intertwiner).
In the same way, toric matrices $W$ of size $d_A^2$ are
replaced by relative toric matrices $W^{rel}$ of size
$d_G^2$.  The modular splitting technique can be applied
as before, with the advantage that the size of tensors is
greatly reduced. Once the relative matrices are found, we
can retrieve the others by the relation $W_{xy} = E_0
{W^{rel}}_{xy} E_0^T$.
Such an example is worked out in the last section of reference
\cite{EstebanRobert}

\subsection{A dual bimodule structure?}

Axioms for quantum groupo\"{\i}ds are certainly self-dual, but the
objects that  we have at hand are not  generic : they are quite
special. In particular, if it is clear that $Oc(G)$ is an $A(G)$
bimodule, there is no obvious reason for $A(G)$ to be an $Oc(G)$
bimodule. If it were so, this action would be defined by a set of
coefficients $P_{xy}$, with $x \, \lambda \, y = \sum_{\mu}
(P_{xy})_{\lambda \mu} \, \mu$. The $P_{xy}$ being of dimension
$d_{{\cal A}_k} \times d_{{\cal A}_k}$ and the bimodule
associativity property $(x x') \lambda (y y') = x (x' \lambda y)
y'$ would  lead to a  double quantum symmetry equation: $ P_{x'y}
\, P_{xy'} = \sum_{x'' y''} \, O_{xx'}^{x''} \, O_{yy'}^{y''}\,
P_{x''y''} $. This equation taken at $y=y'=0$, at $x=x'=0$ and at
$x = y =0$ would itself lead to: $ P_{x'0} \, P_{x0} = \sum_{x''}
O_{xx'}^{x''} \, P_{x'' 0} $ , $ P_{0 y} \, P_{0 y'} = \sum_{y''}
O_{y y'}^{y''} \, P_{0y''} $, $ P_{x'y'} = P_{x'0} \, P_{0y'} =
P_{0y'} \, P_{x'0}$ and each set of matrices $P_{x 0}$ or $P_{0
y}$ would give respectively an anti-representation and a
representation of dimension $d_{{\cal A}_k} \times d_{{\cal A}_k}$
of the quantum symmetry algebra. Now, what could these $P_{xy}$
matrices be? One obvious candidate is to set them equal to the
toric matrices $W_{xy}$. The problem is that this choice cannot
work since, as it can be checked on simple examples,  $W_{x'y'}$
is not equal to $W_{x'0}  W_{0y'}$ in general. Existence of a dual
bimodule structure is not excluded,  but if it exists, it  cannot
be defined by the toric matrices alone. Supposing the existence of
such dual bimodule structure, it should also satisfy some
compatibility conditions, like $(\lambda (x (\mu (y \, a)))) =
((\lambda \, x \, \mu) (y \, a)) = (\lambda (x \, \mu \, y) a))$,
leading to the following set of relations:
\begin{equation}
S_y \, F_{\mu} \, S_x \, F_{\lambda} = \sum_{z} (V_{\lambda \mu})_{xz} \, S_y \, S_z
 = \sum_{\nu} (P_{xy})_{\mu \nu} \, F_{\nu} \, F_{\lambda} \;.
\end{equation}

\subsection{Realization of the Ocneanu quantum symmetries}
In many cases $Oc(G)$ can be written in terms of the tensor square
of the graph algebras of some related graph $K$ with self fusion,
with the tensor product taken over a subalgebra, called the
modular subalgebra $J$. In the simplest cases, \ie when $G$ has
self fusion, $K$ is $G$ itself. The set of elements of $J$ is
determined by modular properties
\cite{Coq-qtetra,GilRobert1,GilRobert2,Gil-thesis}. Each vertex of an
$\mathcal{A}_k$ graph has a fixed modular operator value $T$. The
vector space spanned by vertices of a $G$ graph is a module over
$\mathcal{A}_k$, and one can try to define a modular operator
value on vertices of $G$. Suppose that the vertex $a$ of $G$
appears both in the branching rules (restriction map from
$\mathcal{A}_k$ to $G$) of vertices $\lambda$ and $\mu$ of
$\mathcal{A}_k$. The vertex $a$ will have a well-defined modular
operator value if the two values $T(\lambda)$ and $T(\mu)$ are
equal. The set of vertices having this property is a subalgebra of
the graph algebra of $G$, denoted $J$. 

As already commented, non
trivial multiplicities in the modular matrix lead to non
commutativity for $Oc(G)$. This happens whenever $G$ possesses
classical symmetries\footnote{By this we mean that, the unit
vertex being chosen, the graph still contains a classical symmetry, 
making impossible a direct computation of the table of multiplication.}. 
In those cases, the algebraic realization of $Oc(G)$ involves not only a tensor square
over some subalgebra but a cross product by an appropriate
discrete group algebra \cite{Gil-thesis}. The bimodule structure
of $Oc(G)$ over ${\cal A}_k \otimes {\cal A}_k$ is thus related to
the module structure of $G$ over ${\cal A}_k$.


\section{The $SU(3)$ system of graphs and their quantum symmetries}

\begin{figure}[ttt]
\setlength{\unitlength}{0.6 mm}
\begin{center}
\begin{picture}(180,100)


\put(0,75){\begin{picture}(25,15)\qbezier[50](5,0)(0,6)(5,12)\qbezier[50](5,0)(10,6)(5,12)
\qbezier[50](5,12)(0,18)(5,24)\qbezier[50](5,12)(10,18)(5,24)
\qbezier[40](5,24)(7.5,27)(10,24)\qbezier[40](5,24)(7.5,21)(10,24)
\qbezier[40](5,12)(7.5,15)(10,12)\qbezier[40](5,12)(7.5,9)(10,12)
\put(5,0){\circle*{2}}\put(5,12){\circle*{2}}\put(5,24){\circle*{2}}
\put(3,6){\vector(0,-1){0.5}}\put(7.5,6){\vector(0,1){0}}
\put(3,18){\vector(0,-1){0.5}}\put(7.5,18){\vector(0,1){0}}
\put(-5,32){\makebox(0,0){($\mathcal{A}_{4}^{\ast}$)}}
\end{picture}}


\put(30,72){\begin{picture}(20,40)\put(0,20){\line(1,0){20}}\put(10,20){\vector(1,0){0.5}}
\put(0,20){\line(1,1){10}}\put(5,25){\vector(-1,-1){0.5}}\put(20,20){\line(-1,1){10}}\put(15,25){\vector(-1,1){0.5}}
\put(0,20){\line(2,-1){10}}\put(5,17.5){\vector(2,-1){0.5}}\put(20,20){\line(-2,-1){10}}\put(15,17.5){\vector(2,1){0.5}}
\put(0,20){\line(2,-3){10}}\put(5,12.5){\vector(-2,3){0.5}}\put(20,20){\line(-2,-3){10}}\put(15,12.5){\vector(-2,-3){0.5}}
\qbezier[25](10,5)(6,10)(10,15)\qbezier[25](10,5)(14,10)(10,15)\qbezier[15](10,5)(7,3)(10,1)\qbezier[15](10,5)(13,3)(10,1)
\put(8,10){\vector(0,-1){0}}\put(12,12){\vector(0,1){0}}
\put(10,5){\circle*{2}}\put(10,15){\circle*{2}}\put(0,20){\circle*{2}}\put(20,20){\circle*{2}}\put(10,30){\circle*{2}}
\put(-2,34){\makebox(0,0){($\mathcal{D}_{4}$)}}
\end{picture}}


\put(75,75){\begin{picture}(20,25)\put(0,20){\line(1,0){20}}\put(10,20){\vector(1,0){0.5}}
\put(0,20){\line(2,1){10}}\put(5,22.5){\vector(-2,-1){0.5}}\put(20,20){\line(-2,1){10}}\put(15,22.5){\vector(-2,1){0.5}}
\put(0,20){\line(0,-1){20}}\put(0,15){\vector(0,-1){0}}\put(0,5){\vector(0,1){0}}
\put(20,20){\line(0,-1){20}}\put(20,15){\vector(0,1){0}}\put(20,5){\vector(0,-1){0}}
\put(0,10){\line(2,-1){20}}\put(5,7.5){\vector(-2,1){0}}\put(15,2.5){\vector(-2,1){0}}
\put(20,10){\line(-2,-1){20}}\put(17,8.5){\vector(2,1){0}}\put(3,1.5){\vector(-2,-1){0}}

\put(0,10){\line(1,0){20}}\put(10,10){\vector(-1,0){0.5}}

\put(0,10){\line(1,-1){10}}\put(5,5){\vector(1,-1){0}}\put(20,10){\line(-1,-1){10}}\put(16,6){\vector(1,1){0}}

\put(0,10){\line(2,3){10}}\put(5,17.5){\vector(2,3){0}}\put(20,10){\line(-2,3){10}}\put(15,17.5){\vector(2,-3){0}}

\put(0,20){\line(2,-3){10}}\put(5,12.5){\vector(-2,3){0}}\put(20,20){\line(-2,-3){10}}\put(15,12.5){\vector(-2,-3){0}}

\put(10,25){\circle*{2}}\put(10,5){\circle*{2}}\put(10,0){\circle*{2}}
\put(0,20){\circle*{1.5}}\put(0,20){\circle{3}}\put(20,10){\circle*{1.5}}\put(20,10){\circle{3}}\put(0,0){\circle*{1.5}}\put(0,0){\circle{3}}
\put(20,20){\circle{1.5}}\put(20,20){\circle{3}}\put(0,10){\circle{1.5}}\put(0,10){\circle{3}}\put(20,0){\circle{1.5}}\put(20,0){\circle{3}}
\put(-5,32){\makebox(0,0){($\mathcal{D}_{4}^{\ast}$)}}
\end{picture}}


\put(130,75){\begin{picture}(30,25)\put(0,11){\line(1,0){20}}\put(10,11){\vector(1,0){0}}
\put(0,11){\line(1,1){10}}\put(5,16){\vector(-1,-1){0}}\put(20,11){\line(-1,1){10}}\put(15,16){\vector(-1,1){0}}
\put(0,9){\line(1,0){20}}\put(10,9){\vector(-1,0){0}}
\put(0,9){\line(1,-1){10}}\put(5,4){\vector(1,-1){0}}\put(20,9){\line(-1,-1){10}}\put(15,4){\vector(1,1){0}}
\qbezier[15](-1,10)(-4,12)(-6,10)\qbezier[15](-1,10)(-4,8)(-6,10)
\qbezier[15](21,10)(24,12)(26,10)\qbezier[15](21,10)(24,8)(26,10)
\put(10,-1){\circle*{3}}\put(10,21){\circle*{3}}\put(0,10){\circle*{3}}\put(20,10){\circle*{3}}
\put(-5,30){\makebox(0,0){($\mathcal{E}_{5}^{\ast}$)}}
\end{picture}}


\put(-15,0){\begin{picture}(60,70)
\put(0,10){\line(1,0){60}}\put(0,10){\line(1,2){30}}\put(60,10){\line(-1,2){30}}
\put(0,40){\line(1,0){60}}\put(0,40){\line(3,-4){30}}\put(60,40){\line(-3,-4){30}}
\put(0,40){\line(4,-1){60}}\put(60,40){\line(-4,-1){60}}\put(0,25){\line(4,-1){60}}\put(60,25){\line(-4,-1){60}}
\put(0,10){\line(3,-1){30}}\put(60,10){\line(-3,-1){30}}\put(20,50){\line(1,0){20}}
\put(0,40){\line(2,1){20}}\put(60,40){\line(-2,1){20}}
\put(0,25){\line(4,5){20}}\put(60,25){\line(-4,5){20}}
\put(0,25){\line(1,0){60}}

\put(30,70){\circle*{2}}\put(30,32.5){\circle*{2}}\put(30,17.5){\circle*{2}}\put(30,0){\circle*{2}}
\put(0,40){\circle{1.5}}\put(0,40){\circle{3}}\put(0,25){\circle{1.5}}\put(0,25){\circle{3}}
\put(0,10){\circle{1.5}}\put(0,10){\circle{3}}\put(40,50){\circle{1.5}}\put(40,50){\circle{3}}
\put(60,40){\circle*{1.5}}\put(60,40){\circle{3}}\put(60,25){\circle*{1.5}}\put(60,25){\circle{3}}
\put(60,10){\circle*{1.5}}\put(60,10){\circle{3}}\put(20,50){\circle*{1.5}}\put(20,50){\circle{3}}

\put(25,60){\vector(-1,-2){0}}\put(10,30){\vector(-1,-2){0}}\put(35,60){\vector(-1,2){0}}\put(49,32){\vector(-1,2){0}}
\put(30,10){\vector(-1,0){0}}\put(30,25){\vector(-1,0){0}}\put(30,40){\vector(-1,0){0}}\put(30,50){\vector(1,0){0}}
\put(16.5,17.5){\vector(3,-4){0}}\put(43.5,17.5){\vector(3,4){0}}
\put(10,45){\vector(-2,-1){0}}\put(50,45){\vector(-2,1){0}}\put(15,5){\vector(3,-1){0}}\put(45,5){\vector(3,1){0}}
\put(20,35){\vector(4,-1){0}}\put(40,30){\vector(4,-1){0}}\put(20,30){\vector(4,1){0}}\put(40,35){\vector(4,1){0}}
\put(20,20){\vector(4,-1){0}}\put(40,15){\vector(4,-1){0}}\put(20,15){\vector(4,1){0}}\put(40,20){\vector(4,1){0}}
\put(8,35){\vector(-1,-1){0}}\put(52,35){\vector(-1,1){0}}
\put(0,60){\makebox(0,0){($\mathcal{E}_{9}^{\ast}$)}}
\end{picture}}


\put(50,30){\begin{picture}(90,60)
\put(0,0){\circle*{2}}\put(15,-7.5){\circle*{1.5}}\put(15,-7.5){\circle{3}}\put(30,-15){\circle{1.5}}
\put(30,-15){\circle{3}}\put(60,-30){\circle{1.5}}\put(60,-30){\circle{3}}

\put(30,-30){\circle*{2}}\put(60,-15){\circle*{2}}
\put(75,-7.5){\circle*{1.5}}\put(75,-7.5){\circle{3}}\put(90,0){\circle{1.5}}\put(90,0){\circle{3}}

\put(45,7.5){\circle*{1.5}}\put(45,7.5){\circle{3}}\put(45,22.5){\circle*{1.5}}\put(45,22.5){\circle{3}}
\put(45,-22.5){\circle*{1.5}}\put(45,-22.5){\circle{3}}

\put(30,0){\circle*{2}}\put(15,7.5){\circle{1.5}}\put(15,7.5){\circle{3}}
\put(60,30){\circle*{2}}\put(30,30){\circle{1.5}}\put(30,30){\circle{3}}
\put(75,7.5){\circle*{2}}\put(60,0){\circle{1.5}}\put(60,0){\circle{3}}

\put(0,0){\line(2,-1){15}} \put(0,0){\line(2,1){15}}
\put(15,-7.5){\line(0,1){15}} \put(10,5){\vector(-2,-1){0.5}}
\put(10,-5){\vector(2,-1){0.5}} \put(15,0){\vector(0,1){0.5}}

\put(0,0){\line(2,-1){60}}\put(60,-30){\line(0,1){60}}\put(30,-30){\line(0,1){60}}
\put(30,-30){\line(2,1){60}}\put(90,0){\line(-2,1){60}}\put(0,0){\line(2,1){60}}

\put(15,-7.5){\line(0,1){15}}\put(75,-7.5){\line(0,1){15}}
\put(15,7.5){\line(2,-1){15}}\put(75,7.5){\line(-2,-1){15}}
\put(15,-7.5){\line(2,1){30}}\put(75,-7.5){\line(-2,1){30}}

\put(30,-15){\line(1,0){30}}\put(30,0){\line(1,0){30}}\put(30,30){\line(1,0){30}}
\put(30,-15){\line(2,3){30}}\put(60,-15){\line(-2,3){30}}\put(60,0){\line(-2,-3){15}}\put(30,0){\line(2,-3){15}}
\put(30,0){\line(2,3){15}}\put(60,0){\line(-2,3){15}}

\put(25,2.5){\vector(2,-1){0}} \put(25,-2.5){\vector(2,1){0}}
\put(30,5){\vector(0,-1){0}} \put(25,12.5){\vector(-2,-1){0}}
\put(25,-12.5){\vector(2,-1){0}} \put(30,-5){\vector(0,1){0}}
\put(40,-20){\vector(2,-1){0}} \put(55,-27.5){\vector(2,-1){0}}
\put(40,-25){\vector(2,1){0}} \put(55,-17.5){\vector(-2,-1){0}}
\put(70,-10){\vector(2,1){0}} \put(70,-5){\vector(-2,1){0}}
\put(70,5){\vector(2,1){0}}\put(70,10){\vector(-2,1){0}}
\put(85,-2.5){\vector(2,1){0}} \put(80,5){\vector(-2,1){0}}
\put(40,25){\vector(-2,1){0}} \put(45,30){\vector(1,0){0}}
\put(45,0){\vector(-1,0){0}} \put(45,-15){\vector(1,0){0}}
\put(30,-20){\vector(0,1){0}}\put(60,-20){\vector(0,1){0}}
\put(60,-7.5){\vector(0,-1){0}} \put(60,10){\vector(0,1){0}}
\put(75,0){\vector(0,-1){0}}\put(50,25){\vector(-2,-1){0}}

\put(40,5){\vector(2,1){0}} \put(50,5){\vector(-2,1){0}}
\put(40,15){\vector(-2,-3){0}} \put(50,15){\vector(-2,3){0}}
\put(36,-6){\vector(-2,-3){0}} \put(52,-3){\vector(-2,3){0}}
\put(38,-12){\vector(2,-3){0}} \put(54,-9){\vector(2,3){0}}
\put(10,27){\makebox(0,0){($\mathcal{D}_{9}^{t}$)}}
\end{picture}}


\put(150,5){\begin{picture}(90,60)
\put(10,0){\circle*{2}}\put(10,20){\circle*{2}}\put(10,30){\circle*{2}}
\put(10,51.5){\circle*{2}}\put(10,61.5){\circle*{2}}

\put(5,-5){\circle*{1.5}}\put(5,-5){\circle{3}}\put(15,-5){\circle{1.5}}\put(15,-5){\circle{3}}
\put(0,10){\circle{1.5}}\put(0,10){\circle{3}}\put(20,10){\circle*{1.5}}\put(20,10){\circle{3}}
\put(0,40.75){\circle*{1.5}}\put(0,40.75){\circle{3}}\put(20,40.75){\circle{1.5}}\put(20,40.75){\circle{3}}

\put(0,10){\line(1,0){20}}\put(0,10){\line(1,-1){15}}\put(20,10){\line(-1,-1){15}}
\put(0,10){\line(1,-3){5}}\put(20,10){\line(-1,-3){5}}\put(0,10){\line(0,1){30}}\put(20,10){\line(0,1){30}}
\put(0,10){\line(1,1){10}}\put(0,10){\line(1,2){10}}\put(20,10){\line(-1,1){10}}\put(20,10){\line(-1,2){10}}
\put(0,40){\line(1,-1){10}}\put(0,40){\line(1,-2){10}}\put(20,40){\line(-1,-1){10}}\put(20,40){\line(-1,-2){10}}
\put(0,40){\line(1,0){20}}\put(0,41.5){\line(1,0){20}}
\put(0,41.5){\line(1,1){10}}\put(0,41.5){\line(1,2){10}}\put(20,41.5){\line(-1,1){10}}\put(20,41.5){\line(-1,2){10}}

\put(0,40){\line(1,-4){10}}\put(20,40){\line(-1,-4){10}}

\put(10,41.5){\vector(1,0){0}}\put(10,40){\vector(1,0){0}}\put(5,46.5){\vector(-1,-1){0}}
\put(5,51.5){\vector(-1,-2){0}}\put(5,35){\vector(1,-1){0}}\put(5,30){\vector(1,-2){0}}
\put(15,46.5){\vector(-1,1){0}}\put(15,51.5){\vector(-1,2){0}}\put(15,35){\vector(-1,-1){0}}
\put(15,30){\vector(-1,-2){0}}\put(5,15){\vector(1,1){0}}\put(5,20){\vector(1,2){0}}
\put(15,15){\vector(1,-1){0}}\put(15,20){\vector(1,-2){0}}
\put(7.5,10){\vector(-1,0){0}}\put(2.5,2.5){\vector(-1,3){0}}\put(17.5,2.5){\vector(-1,-3){0}}
\put(5,5){\vector(1,-1){0}}\put(11,-1){\vector(-1,1){0}}
\put(15,5){\vector(1,1){0}}\put(7,-3){\vector(-1,-1){0}}
\put(0,25){\vector(0,-1){0}}\put(20,25){\vector(0,1){0}}
\put(2.5,30){\vector(-1,4){0}}\put(17.5,30){\vector(-1,-4){0}}
\put(-13,53){\makebox(0,0){(${\mathcal{D}_{9}^{t}}^{\ast}$)}}
\end{picture}}

\end{picture}
\end{center}
\caption{Some module graphs without self-fusion: $\mathcal{A}_{4}^{\ast}$,  $%
\mathcal{D}_{4}$, $\mathcal{D}_{4}^{\ast}$, $\mathcal{E}_{5}^{\ast}$, $\mathcal{%
E}_{9}^{\ast}$, $\mathcal{D}_{9}^{t}$, and
${\mathcal{D}_{9}^{t}}^{\ast}$.} 
\label{fig:graphnoself}
\end{figure}
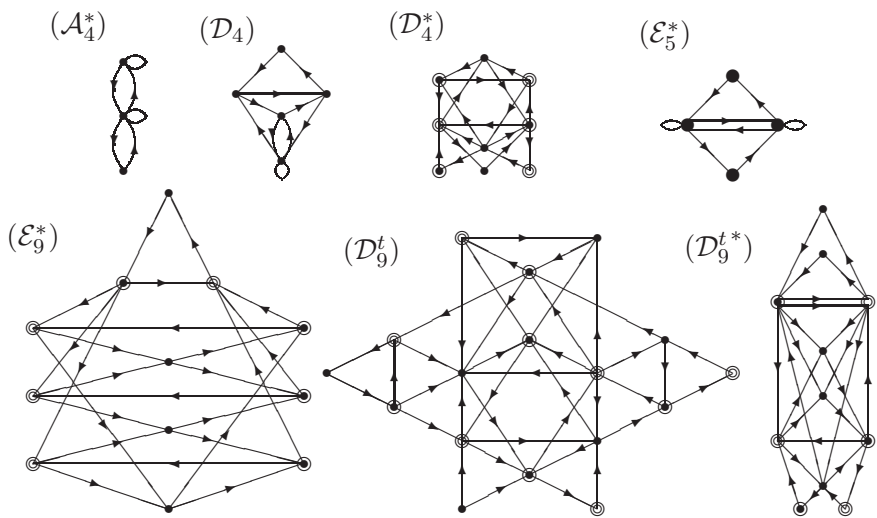

Starting with the complete list of modular invariants
\cite{Gannon-su3}, the list of graphs was found by
\cite{DiFZuber}, slightly amended by \cite{Oc-Bariloche}. We
believe that a determination of the graph of quantum symmetries
associated with the above was worked out in 2000 or before by A.
Ocneanu (unpublished). We now present a compendium of results
concerning not only these quantum symmetries but also several
other results that use the concepts introduced in previous
sections. In particular we give in most cases an algebraic
realization of $Oc(G)$ that allows one to perform calculations
without having to use the graph of quantum symmetries. A detailed
study of several cases has already been made available in the
litterature \cite{GilRobert2,Gil-thesis} and details concerning
the others will be published elsewhere
\cite{GilDahmaneHassan,EstebanGil,Dahmane-thesis}. Several graphs are displayed in figures 
\ref{fig:graphself} and \ref{fig:graphnoself}.

\subsection{The $\mathcal{A}$ series and its conjugated series }

\subsubsection{The $\mathcal{A}$ series (graphs with self-fusion)}

The $\mathcal{A}_k$ graphs are the Weyl alcoves of $SU(3)$ at level $k$. We have $A(\mathcal{A}_k) = \mathcal{A}_k$, so the annular matrices coincide with the fusion matrices: $F_{\lambda} = N_{\lambda}$.
The algebra of quantum symmetries is realized as $Oc(\mathcal{A}_{k}) = \mathcal{A}_{k}\overset{\cdot }{\otimes }\mathcal
{A}_{k} $ where the tensor product is taken over $\mathcal{A}_{k}$ with the
identification $\lambda \overset{\cdot}{\otimes} \mu \equiv
\lambda \mu^{\ast }\overset{\cdot }{\otimes }0$. A basis of $Oc(\mathcal{A}_k)$ is $x=\lambda\otimesdot 0$ and the dimension $d_O = d_{{\mathcal A}_k}$. The dual annular matrices are $S_x = F_{\lambda} = N_{\lambda}$ and the double annular
matrices are $V_{\lambda\mu} = N_{\lambda} N_{\mu^*}$. The modular invariant associated to the $\mathcal{A}_k$ graph is
diagonal $M_{\lambda\mu} = \delta_{\lambda\mu}$. We can easily check that $(V_{\lambda\mu})_{00} = M_{\lambda\mu}$.
The two algebras $\mathcal{BA}_{k}$ and $\widehat{\mathcal{B}}\mathcal
{A}_{k}$ are isomorphic. We have $d_{x}=d_{\lambda}$, the quadratic and linear sum
rules are trivially satisfied.
In the $SU(2)$ system, i.e. for $ADE$ diagrams, the value
of $d_H = \sum_\lambda d_\lambda $ has been
obtained independently, for all diagrams, by A. Ocneanu
(unpublished) and by \cite{Pet_Zub-Occells} and \cite{Coq-qtetra,GilRobert1}.
It is easy to see, for instance,  that for $A_{k+1} = {\cal A}_{k}$
graphs, the following formula holds : $d_H=\frac{(k + 1) (k + 2) (k +
3)}{6}$. Actually, setting $r = k+1$ (the number of vertices) and $
\kappa = k+2$ (the usual Coxeter number), this formula also works for
$D$-graphs and for exceptionals,  when it is written as $d_H = r\kappa
(\kappa +1)/6$. It is interesting to notice that the same formula
also gives the dimension of the Gelfand - Ponomarev preprojective
algebra associated with the chosen graph (see \cite{MOV}).
For the $SU(3)$ system of graphs (now $\kappa = k+3$)  we observe
that the dimension $d_H = \sum_\lambda d_\lambda $ of
graphs  ${\cal A}_{k}$ is given by the formula
\begin{equation}
d_H=\frac{(k + 1) (k + 2) (k + 3) (k + 4) (k + 5) (k^2+ 6 k + 14)}{1680} \;.
\end{equation}

\subsubsection{The $\mathcal{A}^{\ast }$ modules (no self-fusion)}

The $\mathcal{A}_{k}^{\ast}$ graphs are the conjugated graphs of
$\mathcal{A}_k$. Their vertices are the real vertices of
$\mathcal{A}_{k}$ (see for example
\cite{Gab_Gan,Pet_Zub-conj,Evans-conj}).  We have
$A(\mathcal{A}_k^*)=\mathcal{A}_k$. The algebra of quantum
symmetries is realized as $Oc(\mathcal{A}_k^\ast) = \mathcal{A}_k
\overset{\cdot}{\otimes} \mathcal{A}_k$ where the tensor product
is again taken over $\mathcal{A}_k$ but now with the
identification $\lambda \otimesdot \mu \equiv \lambda \mu
\otimesdot 0$. A basis of $Oc(\mathcal{A}_k^*)$ is again
$x=\lambda\otimesdot 0$, and we have $d_O = d_{{\mathcal A}_k}$.
The dual annular matrices are $S_x = F_{\lambda}$ and the
double annular matrices are $V_{\lambda\mu} = N_{\lambda}
N_{\mu}$. The modular invariant is $M_{\lambda\mu} =
\delta_{\lambda\mu^*}$. The two algebras $\mathcal{BA}_{k}$ and
$\widehat{\mathcal{B}}\mathcal {A}_{k}$ are isomorphic. We have
$d_{x}=d_{\lambda}$, the quadratic and linear sum rules are
trivially satisfied.

\subsection{The  $\mathcal{D}$ series  and the conjugated  $\mathcal{D}^*$ series }
The $\mathcal{D}_k = \mathcal{A}_k/3$ graphs are orbifold graphs
of the $\mathcal{A}_k$ graphs. They are obtained from the action
of the geometrical $\mathbb{Z}_{3}$-automorphism $z$ (see
Eq.(\ref{zauto})) on irreps of the $\mathcal{A}_{k}$ graphs
\cite{Kostov,Fendley,DiFZuber}. Vertices of $\mathcal{A}_k$ that
belong to the same orbit lead to a single vertex in the orbifold
graph $\mathcal{D}_k$. When there is a fixed vertex under $z$
(this happens when $k =0 \mod3$), this vertex is triplicated on
the orbifold graph. Among all orbifold graphs $\mathcal{D}_k$, the
$\mathcal{D}_{3n}$ are the only ones that have self-fusion.

\subsubsection{The $\mathcal{D}_{k}$ orbifold modules for $k \not= 0$ mod 3 (no self-fusion)}
For $k\not=0\mod3$, the $\mathcal{D}_k$ graphs have $(k+1)(k+2)/6$
vertices. One can define a graph algebra with non negative integer
structure constants for these graphs, but it is not compatible
with the $\mathcal{A}_{k}$ action. Therefore these graphs don't
have self-fusion. The Ocneanu algebra is realized as
$Oc(\mathcal{D}_{k})=\mathcal{A}_{k} \otimesdot \mathcal{A}_{k}$
where the tensor product is again taken over $\mathcal{A}_{k}$ but
with the identification $\lambda \otimesdot \mu \equiv \lambda
\rho(\mu^*) \otimesdot 0$, where $\rho$ is the Gannon twist (see
Eq.(\ref{gantwist})). A basis of $Oc(\mathcal{D}_k)$ is
$x=\lambda\otimesdot 0$, and we have $d_O = d_{{\mathcal A}_k}$.
The dual annular matrices are $S_x = F_{\lambda}$ and the
double annular matrices are $V_{\lambda\mu} = N_{\lambda}
N_{\rho(\mu^*)}$. The associated modular invariant is
$M_{\lambda\mu} = \delta_{\lambda\rho(\mu)}$. The two algebras
$\mathcal{BD}_{k}$ and $\widehat{\mathcal{B}} \mathcal{D}_{k}$ are
isomorphic. We have $d_{x}=d_{\lambda}$, the quadratic and linear
sum rules are trivially satisfied. The dimensions $d_{\lambda}$ of the
blocks labelled by $\lambda$ (or by $x$, which is the same here)
satisfy $d_{\lambda}(\mathcal{D}_k) = d_{\lambda}(\mathcal{A}_k)/3$. The dimensions
therefore satisfy $\dim(\mathcal{BD}_k) = \dim(\mathcal{BA}_k)/9$.

\subsubsection{The $\mathcal{D}_k^*$ conjugated orbifold modules $k \not= 0$ mod 3 (no self-fusion)}
The conjugated orbifold graphs $\mathcal{D}_{k}^*$ are the
unfolded (i.e. triplicated) graphs of the $\mathcal{A}_{k}^*$ ones
\cite{DiFZuber}, i.e. their adjacency matrices are such that $%
Ad\left( \mathcal{D}_{k}^{\ast }\right) =\sigma _{123}\otimes
Ad\left( \mathcal{A}_{k}^{\ast }\right)$, where 
 $\sigma
_{123}= \tiny \left(
\begin{array}{ccc}
0 & 1 & 0 \\
0 & 0 & 1 \\
1 & 0 & 0
\end{array}
\right)$ \normalsize is the permutation matrix. These graphs are modules over
the fusion algebras $\mathcal{A}_{k}$. The Ocneanu algebra is
realized as $Oc(\mathcal{D}_{k}^*)=\mathcal{A}_{k} \otimesdot
\mathcal{A}_{k}$ where the tensor product is again taken over
$\mathcal{A}_{k}$ but with the identification $\lambda \otimesdot
\mu \equiv \lambda \rho(\mu) \otimesdot 0$, where $\rho$ is the
Gannon twist defined in Eq. (\ref{gantwist}). A basis of $Oc(\mathcal{D}_k^*)$ is again
$x=\lambda\otimesdot 0$, and we have $d_O = d_{{\mathcal A}_k}$.
The dual annular matrices are $S_x = F_{\lambda}$ and the
double annular matrices are $V_{\lambda\mu} = N_{\lambda}
N_{\rho(\mu)}$. The associated modular invariant is
$M_{\lambda\mu} = \delta_{\lambda\rho(\mu^*)}$. The two algebras
$\mathcal{BD}_{k}^*$ and $\widehat{\mathcal{B}} \mathcal{D}_{k}^*$
are isomorphic. We have $d_{x}=d_{\lambda}$, the quadratic and
linear sum rules are trivially satisfied. The dimensions $d_{\lambda}$ of
the blocks labelled by $\lambda$ (or by $x$, which is the same here)
satisfy $d_{\lambda}(\mathcal{D}_k^*) = 3 \, d_{\lambda}(\mathcal{A}_k^*) $. The
dimensions therefore satisfy $\dim(\mathcal{BD}_k^*) = 9 \,
\dim(\mathcal{BA}_k^*)$.

\subsubsection{The $\mathcal{D}_{k}$ orbifolds for $k=0$ mod 3 (self-fusion)}
For $k=0\mod3$, the $\mathcal{A}_k$ graphs have a central vertex
$\mathbb{Z}_3$-invariant, which is triplicated on the orbifold
graph. In this case $\mathcal{D}_{k}$ graphs have
$(\frac{(k+1)(k+2)}{2}-1)/3 + 3$ vertices, and they possess
self-fusion. The associated modular invariant partition function
is:
\begin{equation}
\mathcal{Z}(\mathcal{D}_{k})  =  \frac{1}{3} \sum_{
\lambda | t(\lambda)=0 } |\chi_{\lambda}^k +
\chi_{z(\lambda)}^k+\chi_{z^{2}(\lambda)}^k|^2
\end{equation}
The special vertex $z$-invariant on the $\mathcal{A}_k$ graph
leads to the presence of a coefficient equal to 3 in the modular
invariant. Therefore the algebra of quantum symmetries of
$\mathcal{D}_{3n}$ is non-commutative. A realization is given by a
semi-direct product $Oc(\mathcal{D}_{3n})=\mathcal{D}_{3n} \ltimes
\mathbb{Z}_{3}$ (see \cite{Gilorbi}). The Ocneanu graph is made of
3 copies of the $\mathcal{D}_{3n}$ graph, its dimension is $d_O =
(k+1)(k+2)/2+8$. The quadratic sum rule is satisfied
but the linear sum rule does not hold $d_{H}\neq d_{V}$ (it may be recovered by introducing
appropriate symmetry factors).

\subsubsection{The $\mathcal{D}_{k}^*$ conjugate orbifolds for $k=0$ mod 3 (no self-fusion)}
The conjugate orbifold graphs $\mathcal{D}_{k}^*$ are the unfolded (i.e. triplicated) graphs of the
$\mathcal{A}_{k}^*$ ones \cite{DiFZuber}. These graphs are modules over the fusion algebras $\mathcal{A}_{k}$.
For $k=0\mod3$, the associated modular invariant partition function is
\begin{equation}
\mathcal{Z}(\mathcal{D}_{k}^*) = \frac{1}{3} \sum_{\lambda | t(\lambda)=0 } (\chi_{\lambda}^k + \chi_{z(\lambda)}^k+\chi_{z^2(\lambda)}^k) \, (\overline{\chi_{\lambda^*}^k} + \overline{\chi_{z(\lambda)^*}^k} + \overline{\chi_{z^2(\lambda)^*}^k})
\end{equation}
Its algebra of quantum symmetries is also non-commutative, and can be realized as a conjugated version of
semi-direct product $Oc(\mathcal{D}_{3n})=\mathcal{D}_{3n} \ltimes \mathbb{Z}_{3}$ (see \cite{Gilorbi}). Its dimension is
$d_O(\mathcal{D}_{k}^*) = d_O(\mathcal{D}_{k})$. The quadratic sum rule is satisfied
but the linear sum rule does not hold $d_{H}\neq d_{V}$ (it may be recovered by introducing
appropriate symmetry factors).

\subsection{Exceptional graphs with self-fusion and their modules}

In the $SU(3)$ family, we have three exceptional graphs with
self-fusion, namely $\mathcal{E}_{5}$, $\mathcal{E}_{9}$ and
$\mathcal{E}_{21}$. Diagrams $\mathcal{E}_{5}$ and
$\mathcal{E}_{21}$ are generalizations of the two Dynkin diagrams
$E_6$ and $E_8$. We have also the module graphs
$\mathcal{E}_{5}^{*}=\mathcal{E}_{5}/3$ and
$\mathcal{E}_{9}^{*}=\mathcal{E}_{9}/3$ (they don't have
self-fusion). Finally we have the exceptional graph
$\mathcal{D}_9^t$ obtained from the exceptional twist of the
$\mathcal{D}_{9}$ graph (a generalization of  the $E_7$ Dynkin
diagram), together with the conjugated exceptional graph
${\mathcal{D}_9^t}^{*}$.

\subsubsection{The exceptional $\mathcal{E}_{5}$ graph (self-fusion)}

The $\mathcal{E}_{5}$ graph has self-fusion and has 12 vertices
denoted $1_{i}$ and $2_{j}$ where $i,j=1,2,...,6$. The unit vertex
is $1_0$ and the fundamental conjugated generators are $2_1$ and
$2_2$ (for more details see \cite {GilRobert2} and \cite
{Gil-thesis}). Its quantum mass is $m(\mathcal{E}_5) =
12(2+\sqrt{2})$. The associated modular invariant partition
functions is:
\begin{eqnarray*}
\mathcal{Z}({\mathcal{E}_5}) &=& |\chi_{(0,0)}^5 + \chi_{(2,2)}^5|^2 + |\chi_{(0,2)}^5 + \chi_{(3,2)}^5|^2
 + |\chi_{(2,0)}^5 + \chi_{(2,3)}^5|^2 \\
{ } &+& |\chi_{(2,1)}^5 + \chi_{(0,5)}^5|^2 + |\chi_{(3,0)}^5 + \chi_{(0,3)}^5|^2 + |\chi_{(1,2)}^5 + \chi_{(5,0)}^5|^2 \; .
\end{eqnarray*}
The modular subalgebra is $J=\{1_{i},\;i=1,...,6\}$ and a
realization of the Ocneanu algebra is given by
$Oc(\mathcal{E}_{5})=\mathcal{E}_{5} \otimesdot_{J}
\mathcal{E}_{5}$, with the identifications $a\otimesdot_J u\,b
\equiv a\,u^*\otimesdot_J b$, for all $u\in J$ and
$a,b\in\mathcal{E}_5$. Conjugation on $\mathcal{E}_5$ is defined
as: ${1_{0}}^*={1_{0}}$, ${1_{5}}^*={1_{1}}$, ${1_{4}}^*={1_{2}}$,
${1_{3}}^*={1_{3}}$, ${2_{0}}^*={2_{3}}$, ${2_{1}}^*={2_{2}}$ and
${2_{5}}^*={2_{4}}$ (it corresponds to the symmetry with respect
to the vertical axis joining vertices ${1_{0}}$ and ${1_{3}}$ of
the diagram $\mathcal{E}_5$ given on  Figure
\ref{fig:graphself}). Its dimension is 24 and a basis of
$Oc(\mathcal{E}_{5})$ is given by $a \otimesdot_J 1_0$ and $b
\otimesdot_J 2_0$, for $a,b\in \mathcal{E}_{5}$. The chiral
generators are $2_1 \otimesdot_J 1_0$ and $1_0 \otimesdot_J 2_1
\equiv 1_5 \otimesdot_J 2_0$. The left and right chiral 
subalgebras are $L =\{ a \otimesdot_J 1_0\}$ and $R =\{ 1_0
\otimesdot_J a\}$, and the ambichiral subalgebra is $A =\{ 1_i
\otimesdot_J 1_0 \equiv 1_0 \otimesdot_J 1_i^*\}$.
The quantum mass of $Oc(\mathcal{E}_{5})$ is $m\left[ Oc(\mathcal{E}_{5})\right] =\frac{m\left[ \mathcal{E}_{5}\right] .m%
\left[ \mathcal{E}_{5}\right] }{m\left[ J\right] }=m\left[ \mathcal{A}_{5}%
\right] =48\left( 3+\sqrt{2}\right).$
The linear and quadratic sum rules hold and read $d_H=d_V = 720$, $\dim(\mathcal{BE}_5) = 29\,376$, respectively.

\subsubsection{The exceptional module of the $\mathcal{E}_{5}$ graph (no
self-fusion)}
The $\mathcal{E}_{5}^*=\mathcal{E}_{5}/3$ is the $\mathbb{Z}_3$-orbifold graph of $\mathcal{E}_5$, it has 4 vertices.
It is a module over $\mathcal{A}_{5}$ and
over $\mathcal{E}_{5}$. In particular it has the
same norm $\beta =\left[ 3\right] _{q}=1+\sqrt{2}$ as
$\mathcal{A}_{5}$ and $\mathcal{E}_{5}$. Its quantum mass is $m(\mathcal{E}_5^*) = m(\mathcal{E}_5)/3 = 4(2+\sqrt{2})$.
The associated modular invariant partition function is:
\begin{eqnarray*}
\mathcal{Z}(\mathcal{E}_5^*)  &=& |\chi_{(0,0)}^5 +
\chi_{(2,2)}^5|^2 +
|\chi_{(3,0)}^5 + \chi_{(0,3)}^5|^2  + (\chi_{(0,2)}^5 + \chi_{(3,2)}^5)
(\overline{\chi_{(2,0)}^5}+\overline{\chi_{(2,3)}^5})
\\
{ } &+&
(\chi_{(2,0)}^5 + \chi_{(2,3)}^5)(\overline{\chi_{(0,2)}^5}+\overline{\chi_{(3,2)}^5}) +
(\chi_{(1,2)}^5 + \chi_{(5,0)}^5)(\overline{\chi_{(0,5)}^5}+\overline{\chi_{(2,1)}^5}) \\
{ } &+&  (\chi_{(2,1)}^5 + \chi_{(0,5)}^5)(\overline{\chi_{(1,2)}^5}+
\overline{\chi_{(5,0)}^5}) \; .
\end{eqnarray*}
The Ocneanu algebra is
$Oc(\mathcal{E}_{5}^*)=\mathcal{E}_{5} \, \otimesdot_{J} \, \mathcal{E}_{5}$
where the tensor product is taken over the modular subalgebra $J$
of $\mathcal{E}_5$ but with the identifications $a \, \otimesdot_J \, 
u\,b \equiv a\,u \, \otimesdot_J \, b$, for all $u\in J$ and
$a,b\in\mathcal{E}_5$. The two algebras 
$Oc(\mathcal{E}_5)$ and $Oc(\mathcal{E}_5^*)$ are isomorphic but their realization in terms of tensor products are different. Here the right chiral generator is $1_0 \otimesdot_J 2_1
\equiv 1_1 \otimesdot_J 2_0$. The quantum mass is
$m(Oc(\mathcal{E}_5^*))=m(Oc(\mathcal{E}_5))$. The dimensions of
the blocks labelled by $\lambda$ and $x$ satisfy
$d_{\lambda}(\mathcal{E}_5^*) = d_{\lambda}(\mathcal{E}_5)/3$ and
$d_{x}(\mathcal{E}_5^*) = d_{x}(\mathcal{E}_5)/3$.  The linear and quadratic sum rules hold and read $d_H=d_V = 720/3=240$ and $\dim(\mathcal{BE}_{5}^*)= \dim(\mathcal{BE}_5)/9 = 3\,264$.

\subsubsection{The exceptional $\mathcal{E}_{9}$ graph (self-fusion)}

The $\mathcal{E}_{9}$ graph has self-fusion and possesses $12$
vertices denoted $0_{i},1_{i},2_{i}$ and $3_{i}$ where $i=0,1$
or $2$. Its quantum mass is $m(\mathcal{E}_9) = 36(2+\sqrt{3})$.
The associated modular invariant partition function is:
\begin{equation*}
\mathcal{Z}(\mathcal{E}_9)  =  |\chi_{(0,0)}^9 + \chi_{(0,9)}^9 + \chi_{(9,0)}^9 +
\chi_{(1,4)}^9 + \chi_{(4,1)}^9 + \chi_{(4,4)}^9|^2 + 2 |\chi_{(2,2)}^9 + \chi_{(2,5)}^9 + \chi_{(5,2)}^9|^2
\end{equation*}
The presence of the factor 2 in the second term of the modular invariant indicates that the 
Ocneanu algebra $Oc(\mathcal{E}_9)$ is
non commutative. It is isomorphic to a direct sum of 36
one-dimensional  blocks of $\mathbb{C}$ and of 9 copies of
2-dimensional matrices $M_2(\mathbb{C})$, its dimension is 72. The modular subalgebra
is $J=\{0_{0},1_{0},2_{0}\}$ and the Ocneanu algebra
$Oc(\mathcal{E}_{9})$ involves $\mathcal{E}_{9}
\otimesdot_{J}\mathcal{E}_{9}$ and a non commutative matrix
complement (see \cite{EstebanGil} for more details). The Ocneanu
graph is made of $12 \times 6 =72$ vertices, corresponding to 3
copies of the $\mathcal{E}_9$ graph and 3 copies of its module
graph $\mathcal{E}_9/3$. The quantum mass is
$m(Oc(\mathcal{E}_{9})) = \frac{m(\mathcal{E}_{9}) \,
m(\mathcal{E}_{9})}{m(J)}=m(\mathcal{A}_{9}) =432 (7+4\sqrt{3})$,
where $m(J)=3$. Note that the quadratic sum rule can be checked
($\dim(\mathcal{BE}_9) = \sum_{\lambda}d_{\lambda}^2(\mathcal{E}_9)
 = \sum_x d_x^2(\mathcal{E}_9) = 518\,976$) but the linear sum rule does
not hold: $d_H = 4\,656$ but $d_V=5\,448$.

\subsubsection{The exceptional module of the $\mathcal{E}_{9}$ graph (no
self-fusion)} The $\mathcal{E}_{9}^{\ast }=\mathcal{E}_{9}/3$
graph\ is a module over the graph algebra
$\mathcal{A}_{9}$ and over the graph algebra $\mathcal{E}_{9}.$
It has the same norm $\beta =\left[ 3\right]
_{q}=1+\sqrt{3}$ as $\mathcal{A}_{9}$ and $\mathcal{E}_{9}$. The $\mathcal{E}_9^*$ graph is associated to the same
modular invariant as $\mathcal{E}_9$. Furthermore, the Ocneanu algebra $Oc(\mathcal{E}_9^*)$ is isomorphic
to $Oc(\mathcal{E}_9)$. But the module structures of $\mathcal{E}_9^*$ over $\mathcal{A}_9$ and over
$Oc(\mathcal{E}_9) \equiv Oc(\mathcal{E}_9^*)$ are not the same as for $\mathcal{E}_9$: 
the annular matrices $F_{\lambda}$ and dual annular matrices 
$S_x$ differ from those of $\mathcal{E}_9$. The quadratic sum rule hold and read
$\dim(\mathcal{BE}_9^*) = 754\,272$, but the linear sum rule does not hold: $d_H = 5\,616$ but $d_V=6\,552$.

\subsubsection{The exceptional $\mathcal{E}_{21}$ graph (self-fusion)}
The $\mathcal{E}_{21}$ graph has self-fusion and possesses $24$
vertices denoted $0,2,...,23$. The unit vertex is 0, the
conjugated generators are 1 and 2. Complex conjugation corresponds to the
symmetry with respect to the horizontal axis joining vertices $0$
and $21$ of the $\mathcal{E}_{21}$ graph given on figure
\ref{fig:graphself}. Triality is equal to the labels taken modulo 3. The norm of the 
$\mathcal{E}_{21}$ graph is $\beta
=\frac{1}{2}({1+\sqrt{2}+\sqrt{6}})$. Actually all quantum dimensions are of the kind 
$(a,b,c,d) = a + b \sqrt{2} + c\sqrt{3} + d\sqrt{6}$, for appropriate values of $a,b,c,d$. The quantum mass is
$m(\mathcal{E}_{21}) = 24(18+10\sqrt{3}+\sqrt{6(97+56\sqrt{3})})$.
The associated modular invariant partition function is:
\begin{eqnarray*}
\mathcal{Z}(\mathcal{E}_{21}) &=&
|\chi_{(0,0)}^{21} + \chi_{(4,4)}^{21} + \chi_{(6,6)}^{21} + \chi_{(10,10)}^{21}+
 \chi_{(0,21)}^{21} + \chi_{(21,0)}^{21} +  \chi_{(1,10)}^{21} + \chi_{(10,1)}^{21} + \chi_{(4,13)}^{21} \\
{} &+& \chi_{(13,4)}^{21} + \chi_{(6,9)}^{21} + \chi_{(9,6)}^{21}|^2  + |\chi_{(0,6)}^{21} + \chi_{(6,0)}^{21} + \chi_{(0,15)}^{21} + \chi_{(15,0)}^{21}+
 \chi_{(4,7)}^{21} + \chi_{(7,4)}^{21} \\
{} &+&  \chi_{(4,10)}^{21} + \chi_{(10,4)}^{21} + \chi_{(6,15)}^{21} + \chi_{(15,6)}^{21}+
 \chi_{(7,10)}^{21} + \chi_{(10,7)}^{21}|^2
\end{eqnarray*}
The modular subalgebra is $J=\{0,21\}$ and a realization of the
Ocneanu algebra is
$Oc(\mathcal{E}_{21})=\mathcal{E}_{21}\otimesdot_{J}
\mathcal{E}_{21}$, with the identifications $a\otimesdot_J u\,b
\equiv a\,u^*\otimesdot_J b$, for all $u\in J$ and
$a,b\in\mathcal{E}_{21}$.  The Ocneanu graph involves $12$
copies of $\mathcal{E}_{21}$. The dimension of
$Oc(\mathcal{E}_{21})$ is 288 (see \cite{GilRobert2, Gil-thesis}
for more details). Its quantum mass is given by $m\left[
Oc(\mathcal{E}_{21})\right] =\frac{m\left[
\mathcal{E}_{21}\right]\, m\left[ \mathcal{E}_{21}\right]
}{m\left[ J\right] }=m\left[ \mathcal{A}_{21} \right],$ where
$m\left[ J\right]=2.$ Numerically $m\left[
Oc(\mathcal{E}_{21})\right]
=1728(201+142\sqrt{2}+116\sqrt{3}+82\sqrt{6}).$ The linear and
quadratic sum rules hold and read $d_H=d_V = 288\,576$,
$\dim(\mathcal{BE}_{21}) = 480\,701\,952$, respectively.

\subsubsection{The twisted exceptional $D_{9}^{t}$ (no self-fusion)}
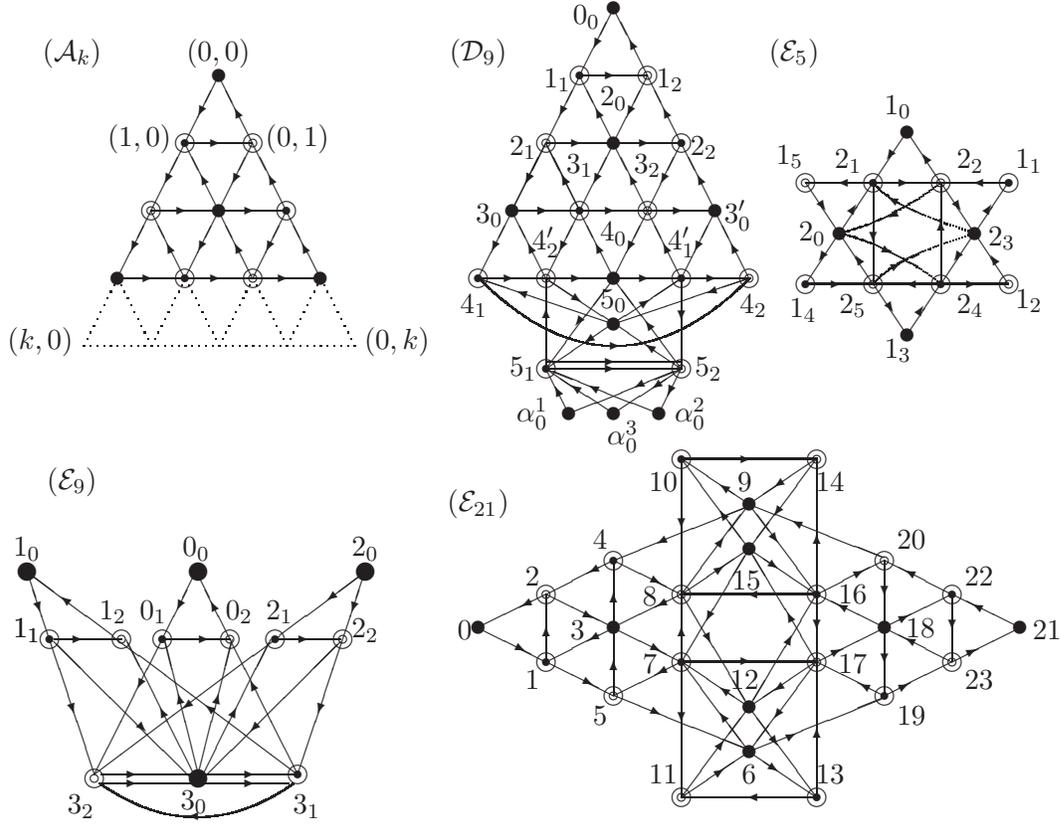
\begin{figure}[ttt]
\setlength{\unitlength}{0.6 mm}
\begin{center}
\begin{picture}(250,190)


\put(20,110){\begin{picture}(15,15) \put(0,0){\line(1,0){15}}
\put(0,0){\line(1,2){7.5}} \put(15,0){\line(-1,2){7.5}}
\put(7.5,0){\vector(1,0){0.5}} \put(11.5,7){\vector(-1,2){0,5}}
\put(3.5,7){\vector(-1,-2){0,5}}
\end{picture}}

\put(35,110){\begin{picture}(15,15) \put(0,0){\line(1,0){15}}
\put(0,0){\line(1,2){7.5}} \put(15,0){\line(-1,2){7.5}}
\put(7.5,0){\vector(1,0){0.5}} \put(11.5,7){\vector(-1,2){0,5}}
\put(3.5,7){\vector(-1,-2){0,5}}
\end{picture}}

\put(50,110){\begin{picture}(15,15) \put(0,0){\line(1,0){15}}
\put(0,0){\line(1,2){7.5}} \put(15,0){\line(-1,2){7.5}}
\put(7.5,0){\vector(1,0){0.5}} \put(11.5,7){\vector(-1,2){0,5}}
\put(3.5,7){\vector(-1,-2){0,5}}
\end{picture}}

\put(27.5,125){\begin{picture}(15,15) \put(0,0){\line(1,0){15}}
\put(0,0){\line(1,2){7.5}} \put(15,0){\line(-1,2){7.5}}
\put(7.5,0){\vector(1,0){0.5}} \put(11.5,7){\vector(-1,2){0,5}}
\put(3.5,7){\vector(-1,-2){0,5}}
\end{picture}}

\put(42.5,125){\begin{picture}(15,15) \put(0,0){\line(1,0){15}}
\put(0,0){\line(1,2){7.5}} \put(15,0){\line(-1,2){7.5}}
\put(7.5,0){\vector(1,0){0.5}} \put(11.5,7){\vector(-1,2){0,5}}
\put(3.5,7){\vector(-1,-2){0,5}}
\end{picture}}

\put(35,140){\begin{picture}(15,15) \put(0,0){\line(1,0){15}}
\put(0,0){\line(1,2){7.5}} \put(15,0){\line(-1,2){7.5}}
\put(7.5,0){\vector(1,0){0.5}} \put(11.5,7){\vector(-1,2){0,5}}
\put(3.5,7){\vector(-1,-2){0,5}}
\end{picture}}

\bezier{40}(12.5,95),(42.5,95),(72.5,95)\bezier{10}(12.5,95),(16.25,102.5),(20,110)
\bezier{10}(20,110),(23.75,102.5),(27.5,95)\bezier{10}(27.5,95),(31.25,102.5),(35,110)
\bezier{10}(35,110),(38.75,102.5),(42.5,95)\bezier{10}(42.5,95),(46.25,102.5),(50,110)
\bezier{10}(50,110),(53.75,102.5),(57.5,95)\bezier{10}(57.5,95),(61.25,102.5),(65,110)
\bezier{10}(65,110),(68.75,102.5),(72.5,95)

\put(20,110){\circle*{3}} \put(65,110){\circle*{3}}
\put(42.5,125){\circle*{3}} \put(42.5,155){\circle*{3}}
\put(35,110){\circle*{2}} \put(35,110){\circle{4}}
\put(57.5,125){\circle*{2}} \put(57.5,125){\circle{4}}
\put(35,140){\circle*{2}} \put(35,140){\circle{4}}
\put(50,110){\circle{2}} \put(50,110){\circle{4}}
\put(27.5,125){\circle{2}} \put(27.5,125){\circle{4}}
\put(50,140){\circle{2}} \put(50,140){\circle{4}}

\put(25,141){\makebox(0,0){$(1,0)$}}
\put(60,141){\makebox(0,0){$(0,1)$}}
\put(42.5,160){\makebox(0,0){$(0,0)$}}
\put(82,96){\makebox(0,0){$(0,k)$}}
\put(3,96){\makebox(0,0){$(k,0)$}}

\put(0,32){\makebox(0,0){$1_1$}}

\put(10,160){\makebox(0,0){($\mathcal{A}_{k}$)}}


\put(100,110){\begin{picture}(15,15) \put(0,0){\line(1,0){15}}
\put(0,0){\line(1,2){7.5}} \put(15,0){\line(-1,2){7.5}}
\put(7.5,0){\vector(1,0){0.5}} \put(11.5,7){\vector(-1,2){0,5}}
\put(3.5,7){\vector(-1,-2){0,5}}
\end{picture}}

\put(115,110){\begin{picture}(15,15) \put(0,0){\line(1,0){15}}
\put(0,0){\line(1,2){7.5}} \put(15,0){\line(-1,2){7.5}}
\put(7.5,0){\vector(1,0){0.5}} \put(11.5,7){\vector(-1,2){0,5}}
\put(3.5,7){\vector(-1,-2){0,5}}
\end{picture}}

\put(130,110){\begin{picture}(15,15) \put(0,0){\line(1,0){15}}
\put(0,0){\line(1,2){7.5}} \put(15,0){\line(-1,2){7.5}}
\put(7.5,0){\vector(1,0){0.5}} \put(11.5,7){\vector(-1,2){0,5}}
\put(3.5,7){\vector(-1,-2){0,5}}
\end{picture}}

\put(145,110){\begin{picture}(15,15) \put(0,0){\line(1,0){15}}
\put(0,0){\line(1,2){7.5}} \put(15,0){\line(-1,2){7.5}}
\put(7.5,0){\vector(1,0){0.5}} \put(11.5,7){\vector(-1,2){0,5}}
\put(3.5,7){\vector(-1,-2){0,5}}
\end{picture}}

\put(107.5,125){\begin{picture}(15,15) \put(0,0){\line(1,0){15}}
\put(0,0){\line(1,2){7.5}} \put(15,0){\line(-1,2){7.5}}
\put(7.5,0){\vector(1,0){0.5}} \put(11.5,7){\vector(-1,2){0,5}}
\put(3.5,7){\vector(-1,-2){0,5}}
\end{picture}}

\put(107.5,125){\begin{picture}(15,15) \put(0,0){\line(1,0){15}}
\put(0,0){\line(1,2){7.5}} \put(15,0){\line(-1,2){7.5}}
\put(7.5,0){\vector(1,0){0.5}} \put(11.5,7){\vector(-1,2){0,5}}
\put(3.5,7){\vector(-1,-2){0,5}}
\end{picture}}

\put(122.5,125){\begin{picture}(15,15) \put(0,0){\line(1,0){15}}
\put(0,0){\line(1,2){7.5}} \put(15,0){\line(-1,2){7.5}}
\put(7.5,0){\vector(1,0){0.5}} \put(11.5,7){\vector(-1,2){0,5}}
\put(3.5,7){\vector(-1,-2){0,5}}
\end{picture}}

\put(137.5,125){\begin{picture}(15,15) \put(0,0){\line(1,0){15}}
\put(0,0){\line(1,2){7.5}} \put(15,0){\line(-1,2){7.5}}
\put(7.5,0){\vector(1,0){0.5}} \put(11.5,7){\vector(-1,2){0,5}}
\put(3.5,7){\vector(-1,-2){0,5}}
\end{picture}}

\put(115,140){\begin{picture}(15,15) \put(0,0){\line(1,0){15}}
\put(0,0){\line(1,2){7.5}} \put(15,0){\line(-1,2){7.5}}
\put(7.5,0){\vector(1,0){0.5}} \put(11.5,7){\vector(-1,2){0,5}}
\put(3.5,7){\vector(-1,-2){0,5}}
\end{picture}}

\put(130,140){\begin{picture}(15,15) \put(0,0){\line(1,0){15}}
\put(0,0){\line(1,2){7.5}} \put(15,0){\line(-1,2){7.5}}
\put(7.5,0){\vector(1,0){0.5}} \put(11.5,7){\vector(-1,2){0,5}}
\put(3.5,7){\vector(-1,-2){0,5}}
\end{picture}}

\put(122.5,155){\begin{picture}(15,15) \put(0,0){\line(1,0){15}}
\put(0,0){\line(1,2){7.5}} \put(15,0){\line(-1,2){7.5}}
\put(7.5,0){\vector(1,0){0.5}} \put(11.5,7){\vector(-1,2){0,5}}
\put(3.5,7){\vector(-1,-2){0,5}}
\end{picture}}

\put(115,103){\vector(0,1){0.5}} \put(115,110){\line(0,-1){20}}
\put(145,101){\vector(0,-1){0.5}} \put(145,110){\line(0,-1){20}}
\put(130,91.5){\vector(1,0){0.5}} \put(115,91.5){\line(1,0){30}}
\put(130,90){\vector(1,0){0.5}} \put(115,90){\line(1,0){30}}

\put(135,86){\vector(-2,-1){0.5}} \put(120,80){\line(5,2){25}}
\put(125,86){\vector(-2,1){0.5}} \put(140,80){\line(-5,2){25}}
\put(117.5,85){\vector(-1,2){0.5}} \put(120,80){\line(-1,2){5}}
\put(142.5,85){\vector(-1,-2){0.5}} \put(140,80){\line(1,2){5}}
\put(139,86){\vector(-3,-2){0.5}} \put(130,80){\line(3,2){15}}
\put(121,86){\vector(-3,2){0.5}} \put(130,80){\line(-3,2){15}}

 \put(115,90){\line(3,2){30}}\put(124,96){\vector(-3,-2){0.5}}\put(137.5,95){\vector(-3,2){0.5}}
 \put(145,90){\line(-3,2){30}}\put(122.5,105){\vector(3,-2){0.5}}\put(137.5,105){\vector(3,2){0.5}}

\put(115,90){\line(3,4){15}}\put(122,100){\vector(-3,-4){0.5}}
\put(145,90){\line(-3,4){15}}\put(137.5,100){\vector(-3,4){0.5}}

\put(130,100){\line(3,1){30}}\put(151,107){\vector(-3,-1){0.5}}
\put(130,100){\line(-3,1){30}}\put(109,107){\vector(-3,1){0.5}}

\qbezier[500](100,110)(130,80)(160,110)\put(131,95){\vector(1,0){0.5}}

\put(124,168){\makebox(0,0){$0_0$}}
\put(117.5,154){\makebox(0,0){$1_1$}}
\put(142.5,154){\makebox(0,0){$1_2$}}
\put(110,139){\makebox(0,0){$2_1$}}
\put(150,139){\makebox(0,0){$2_2$}}
\put(102.5,124){\makebox(0,0){$3_0$}}
\put(157.5,124){\makebox(0,0){$3_0^{\prime}$}}
\put(99,104){\makebox(0,0){$4_1$}}
\put(161,104){\makebox(0,0){$4_2$}}
\put(130,150){\makebox(0,0){$2_0$}}
\put(122.5,135){\makebox(0,0){$3_1$}}
\put(137.5,135){\makebox(0,0){$3_2$}}
\put(130,120){\makebox(0,0){$4_0$}}
\put(115,118){\makebox(0,0){$4_2^{\prime}$}}
\put(145,118){\makebox(0,0){$4_1^{\prime}$}}
\put(130,105){\makebox(0,0){$5_0$}}
\put(110,90){\makebox(0,0){$5_1$}}
\put(151,90){\makebox(0,0){$5_2$}}
\put(112,80){\makebox(0,0){$\alpha^{1}_0$}}
\put(147,80){\makebox(0,0){$\alpha^{2}_0$}}
\put(132,75){\makebox(0,0){$\alpha^{3}_0$}}

\put(130,170){\circle*{3}} \put(130,140){\circle*{3}}
\put(130,110){\circle*{3}} \put(107.5,125){\circle*{3}}
\put(152.5,125){\circle*{3}}

\put(122.5,155){\circle*{2}} \put(122.5,155){\circle{4}}
\put(145,140){\circle*{2}} \put(145,140){\circle{4}}
\put(122.5,125){\circle*{2}} \put(122.5,125){\circle{4}}
\put(100,110){\circle*{2}} \put(100,110){\circle{4}}
\put(145,110){\circle*{2}} \put(145,110){\circle{4}}

\put(137.5,155){\circle{2}} \put(137.5,155){\circle{4}}
\put(115,140){\circle{2}} \put(115,140){\circle{4}}
\put(137.5,125){\circle{2}} \put(137.5,125){\circle{4}}
\put(115,110){\circle{2}} \put(115,110){\circle{4}}
\put(160,110){\circle{2}} \put(160,110){\circle{4}}

\put(120,80){\circle*{3}} \put(140,80){\circle*{3}}
\put(130,80){\circle*{3}} \put(130,100){\circle*{3}}
\put(115,90){\circle*{2}} \put(115,90){\circle{4}}
\put(145,90){\circle{2}} \put(145,90){\circle{4}}

\put(100,160){\makebox(0,0){($\mathcal{D}_9$)}}


\put(172.5,108.75){\begin{picture}(15,11.25)
\put(0,0){\line(1,0){15}} \put(0,0){\line(2,3){7.5}}
\put(7.5,11.25){\line(2,-3){7.5}} \put(7.5,0){\vector(1,0){0.5}}
\put(11.5,5.75){\vector(-2,3){0.5}}
\put(3.5,5.75){\vector(-2,-3){0.5}}
\end{picture}}

\put(202.5,108.75){\begin{picture}(15,11.25)
\put(0,0){\line(1,0){15}} \put(0,0){\line(2,3){7.5}}
\put(7.5,11.25){\line(2,-3){7.5}} \put(7.5,0){\vector(1,0){0.5}}
\put(11.5,5.75){\vector(-2,3){0.5}}
\put(3.5,5.75){\vector(-2,-3){0.5}}
\end{picture}}

\put(187.5,131.25){\begin{picture}(15,11.25)
\put(0,0){\line(1,0){15}} \put(0,0){\line(2,3){7.5}}
\put(7.5,11.25){\line(2,-3){7.5}} \put(7.5,0){\vector(1,0){0.5}}
\put(11.5,5.75){\vector(-2,3){0.5}}
\put(3.5,5.75){\vector(-2,-3){0.5}}
\end{picture}}

\put(180,120){\begin{picture}(15,11.25)
\put(0,0){\line(-2,3){7.5}} \put(-7.5,11.25){\line(1,0){15}}
\put(7.5,11.25){\line(-2,-3){7.5}}
\put(0,11.25){\vector(-1,0){0.5}}
\put(3.25,5.75){\vector(2,3){0.5}}
\put(-3.5,5.75){\vector(2,-3){0.5}}
\end{picture}}

\put(195,97.5){\begin{picture}(15,11.25)
\put(0,0){\line(-2,3){7.5}} \put(-7.5,11.25){\line(1,0){15}}
\put(7.5,11.25){\line(-2,-3){7.5}}
\put(0,11.25){\vector(-1,0){0.5}}
\put(3.25,5.75){\vector(2,3){0.5}}
\put(-3.5,5.75){\vector(2,-3){0.5}}
\end{picture}}

\put(210,120){\begin{picture}(15,11.25)
\put(0,0){\line(-2,3){7.5}} \put(-7.5,11.25){\line(1,0){15}}
\put(7.5,11.25){\line(-2,-3){7.5}}
\put(0,11.25){\vector(-1,0){0.5}}
\put(3.25,5.75){\vector(2,3){0.5}}
\put(-3.5,5.75){\vector(2,-3){0.5}}
\end{picture}}

\bezier{50}(180,120),(191.25,122.75),(202.5,131.25)\bezier{70}(202.5,131.25),(202.5,119.25),(202.5,108.75)
\bezier{50}(180,120),(191.25,117.25),(202.5,108.75)

\put(202.5,122.5){\vector(0,1){0.5}}
\put(191.25,123.75){\vector(-2,-1){0.5}}
\put(191.25,115.75){\vector(2,-1){0.5}}

\bezier{50}(187.5,131.25),(191.25,125.75),(210,120)\bezier{50}(210,120),(191.25,114.25),(187.5,108.75)
\bezier{70}(187.5,131.25),(187.5,119.25),(187.5,108.75)

\put(187.5,118.5){\vector(0,-1){0.5}}
\put(191.25,127.75){\vector(-2,1){0.5}}
\put(193.25,113.25){\vector(2,1){0.5}}

 \put(180,120){\circle*{3}}\put(210,120){\circle*{3}} \put(195,142.5){\circle*{3}}\put(195,97.5){\circle*{3}}
 \put(172.5,131.25){\circle{4}}\put(172.5,131.25){\circle{2}}\put(202.5,131.25){\circle{4}}\put(202.5,131.25){\circle{2}}
 \put(187.5,108.75){\circle{4}}\put(187.5,108.75){\circle{2}}\put(217.5,108.75){\circle{4}}\put(217.5,108.75){\circle{2}}
 \put(172.5,108.75){\circle{4}}\put(172.5,108.75){\circle*{2}}\put(202.5,108.75){\circle{4}}\put(202.5,108.75){\circle*{2}}
 \put(187.5,131.25){\circle{4}}\put(187.5,131.25){\circle*{2}} \put(217.5,131.25){\circle{4}}\put(217.5,131.25){\circle*{2}}

\put(172,103){\makebox(0,0){$1_4$}}
\put(222,105){\makebox(0,0){$1_2$}}
\put(169,137){\makebox(0,0){$1_5$}}
\put(222,136){\makebox(0,0){$1_1$}}
\put(193,94){\makebox(0,0){$1_3$}}
\put(193,148){\makebox(0,0){$1_0$}}
\put(183,104){\makebox(0,0){$2_5$}}
\put(209,104){\makebox(0,0){$2_4$}}
\put(182,136){\makebox(0,0){$2_1$}}
\put(209,136){\makebox(0,0){$2_2$}}
\put(174,120){\makebox(0,0){$2_0$}}
\put(216,119){\makebox(0,0){$2_3$}}

\put(170,160){\makebox(0,0){($\mathcal{E}_{5}$)}}


\put(15,0){\line(-1,3){15}}\put(60,0){\line(1,3){15}}
\put(15,0){\line(1,2){22.5}}\put(60,0){\line(-1,2){22.5}}

\put(16,-2){\line(1,0){21}}\put(16.5,0){\line(1,0){21.5}}
\put(39,-2){\line(1,0){19}}\put(39,0){\line(1,0){21.5}}

\put(6,30){\line(1,0){15}}\put(69.5,30){\line(-1,0){14}}\put(30,30){\line(1,0){15}}

\put(38,0){\line(-1,4){7.5}}\put(38,0){\line(1,4){7.5}}

\put(37,0){\line(-1,2){15}}\put(39,0){\line(1,2){15}}

\put(36,-0.5){\line(-1,1){30}}\put(39.5,-0.5){\line(1,1){30}}

\put(15,0){\line(4,3){60}}\put(60,0){\line(-4,3){60}}

\qbezier[500](16,-2)(37.5,-17)(59,-2)

\put(24,0){\vector(1,0){0.5}} \put(50,0){\vector(1,0){0.5}}
\put(24,-2){\vector(1,0){0.5}}\put(50,-2){\vector(1,0){0.5}}
\put(12,30){\vector(1,0){0.5}} \put(62,30){\vector(1,0){0.5}}
\put(40,30){\vector(1,0){0.5}} \put(36,-9.25){\vector(-1,0){0.5}}

\put(33.75,37.5){\vector(-1,-2){0.5}}\put(25,20){\vector(-1,-2){0.5}}
\put(41.25,37.5){\vector(-1,2){0.5}}\put(50,20){\vector(-1,2){0.5}}

\put(10,15){\vector(1,-3){0.5}}\put(2.5,37.5){\vector(1,-3){0.5}}
\put(65,15){\vector(-1,-3){0.5}}\put(72.5,37.5){\vector(-1,-3){0.5}}

\put(35,15){\vector(-4,-3){0.5}}\put(67,39){\vector(-4,-3){0.5}}
\put(40,15){\vector(-4,3){0.5}}\put(8,39){\vector(-4,3){0.5}}

\put(42,16){\vector(1,4){0.5}}\put(34,16){\vector(-1,4){0.5}}

\put(26,10){\vector(1,-1){0.5}}\put(50,10){\vector(-1,-1){0.5}}
\put(30,14){\vector(-1,2){0.5}}\put(46,14){\vector(1,2){0.5}}

\put(0,45){\circle*{4}}\put(38,45){\circle*{4}}
\put(38,-1){\circle*{4}}\put(75,45){\circle*{4}}

\put(5,30){\circle{4}}\put(5,30){\circle*{2}}
\put(30,30){\circle{4}}\put(30,30){\circle*{2}}
\put(55,30){\circle{4}}\put(55,30){\circle*{2}}
\put(60,0){\circle{4}}\put(60,0){\circle*{2}}

\put(21,30){\circle{4}}\put(21,30){\circle{2}}
\put(45,30){\circle{4}}\put(45,30){\circle{2}}
\put(70,30){\circle{4}}\put(70,30){\circle{2}}
\put(15,-1){\circle{4}}\put(15,-1){\circle{2}}

\put(0,50){\makebox(0,0){$1_0$}}
\put(37.5,50){\makebox(0,0){$0_0$}}
\put(75,50){\makebox(0,0){$2_0$}}

\put(0,32){\makebox(0,0){$1_1$}} \put(28,36){\makebox(0,0){$0_1$}}
\put(47,36){\makebox(0,0){$0_2$}}

\put(19,36){\makebox(0,0){$1_2$}}
\put(56,36){\makebox(0,0){$2_1$}}
\put(75,32){\makebox(0,0){$2_2$}}

\put(12,-7){\makebox(0,0){$3_2$}}
\put(37,-6){\makebox(0,0){$3_0$}}
\put(62,-7){\makebox(0,0){$3_1$}}

\put(10,65){\makebox(0,0){($\mathcal{%
E}_{9}$)}}


\put(100,32.5){\begin{picture}(15,15) \put(0,0){\circle*{3}}
\put(15,-7.5){\circle*{2}} \put(15,-7.5){\circle{4}}
\put(15,7.5){\circle{2}} \put(15,7.5){\circle{4}}
\put(0,0){\line(2,-1){15}} \put(0,0){\line(2,1){15}}
\put(15,-7.5){\line(0,1){15}} \put(10,5){\vector(-2,-1){0.5}}
\put(10,-5){\vector(2,-1){0.5}} \put(15,0){\vector(0,1){0.5}}
\end{picture}}

\put(115,25){\begin{picture}(15,15) \put(15,-7.5){\circle{2}}
\put(15,-7.5){\circle{4}}\put(0,0){\line(2,-1){15}}
\put(0,0){\line(2,1){15}} \put(15,-7.5){\line(0,1){15}}
\put(10,5){\vector(-2,-1){0.5}} \put(10,-5){\vector(2,-1){0.5}}
\put(15,0){\vector(0,1){0.5}}
\end{picture}}

\put(115,40){\begin{picture}(15,15) \put(15,7.5){\circle*{2}}
\put(15,7.5){\circle{4}} \put(0,0){\line(2,-1){15}}
\put(0,0){\line(2,1){15}} \put(15,-7.5){\line(0,1){15}}
\put(10,5){\vector(-2,-1){0.5}} \put(10,-5){\vector(2,-1){0.5}}
\put(15,0){\vector(0,1){0.5}}
\end{picture}}

\put(130,32.5){\begin{picture}(15,15) \put(0,0){\circle*{3}}
\put(15,-7.5){\circle*{2}} \put(15,-7.5){\circle{4}}
\put(15,7.5){\circle{2}} \put(15,7.5){\circle{4}}
\put(0,0){\line(2,-1){15}} \put(0,0){\line(2,1){15}}
\put(15,-7.5){\line(0,1){15}} \put(10,5){\vector(-2,-1){0.5}}
\put(10,-5){\vector(2,-1){0.5}} \put(15,0){\vector(0,1){0.5}}
\end{picture}}

\put(130,47.5){\begin{picture}(15,15) \put(0,0){\line(2,-1){15}}
\put(10,-5){\vector(2,-1){0.5}}
\end{picture}}

\put(130,17.5){\begin{picture}(15,15) \put(0,0){\line(2,1){15}}
\put(10,5){\vector(-2,-1){0.5}}
\end{picture}}


\put(205,32.5){\begin{picture}(15,15) \put(15,0){\circle*{3}}
\put(0,7.5){\circle*{2}} \put(0,7.5){\circle{4}}
\put(0,-7.5){\circle{2}} \put(0,-7.5){\circle{4}}
\put(15,0){\line(-2,-1){15}} \put(15,0){\line(-2,1){15}}
\put(0,-7.5){\line(0,1){15}} \put(5,5){\vector(-2,1){0.5}}
\put(5,-5){\vector(2,1){0.5}} \put(0,0){\vector(0,-1){0.5}}
\end{picture}}

\put(190,25){\begin{picture}(15,15) \put(0,-7.5){\circle*{2}}
\put(0,-7.5){\circle{4}}  \put(15,0){\line(-2,-1){15}}
\put(15,0){\line(-2,1){15}} \put(0,-7.5){\line(0,1){15}}
\put(5,5){\vector(-2,1){0.5}} \put(5,-5){\vector(2,1){0.5}}
\put(0,0){\vector(0,-1){0.5}}
\end{picture}}

\put(190,40){\begin{picture}(15,15) \put(0,7.5){\circle{2}}
\put(0,7.5){\circle{4}} \put(15,0){\line(-2,-1){15}}
\put(15,0){\line(-2,1){15}} \put(0,-7.5){\line(0,1){15}}
\put(5,5){\vector(-2,1){0.5}} \put(5,-5){\vector(2,1){0.5}}
\put(0,0){\vector(0,-1){0.5}}
\end{picture}}

\put(175,32.5){\begin{picture}(15,15) \put(15,0){\circle*{3}}
\put(0,7.5){\circle*{2}} \put(0,7.5){\circle{4}}
\put(0,-7.5){\circle{2}} \put(0,-7.5){\circle{4}}
\put(15,0){\line(-2,-1){15}} \put(15,0){\line(-2,1){15}}
\put(0,-7.5){\line(0,1){15}} \put(5,5){\vector(-2,1){0.5}}
\put(5,-5){\vector(2,1){0.5}} \put(0,0){\vector(0,-1){0.5}}
\end{picture}}

\put(175,40){\begin{picture}(15,15) \put(0,0){\line(2,1){15}}
\put(10,5){\vector(2,1){0.5}}
\end{picture}}

\put(175,25){\begin{picture}(15,15) \put(0,0){\line(2,-1){15}}
\put(10,-5){\vector(-2,1){0.5}}
\end{picture}}


\put(145,-5){\begin{picture}(30,30) \put(0,0){\circle{2}}
\put(0,0){\circle{4}} \put(30,0){\circle{4}}
\put(30,0){\circle*{2}}\put(15,10){\circle*{3}}
\put(15,20){\circle*{3}}
\put(0,0){\line(1,0){30}}\put(0,0){\line(0,1){30}}
\put(30,0){\line(0,1){30}} \put(30,30){\line(-1,0){30}}
\put(15,0){\vector(-1,0){0.5}} \put(0,17){\vector(0,-1){0.5}}
\put(30,15){\vector(0,1){0.5}}\put(15,30){\vector(1,0){0.5}}
\put(0,0){\line(3,4){15}}\put(9,12){\vector(3,4){0.5}}
\put(30,0){\line(-3,4){15}}\put(21,12){\vector(3,-4){0.5}}
\put(0,0){\line(3,2){15}}\put(9,6){\vector(3,2){0.5}}
\put(30,0){\line(-3,2){15}}\put(21,6){\vector(3,-2){0.5}}
\put(0,30){\line(3,-4){15}}\put(9,18){\vector(-3,4){0.5}}
\put(0,30){\line(3,-2){15}}\put(9,24){\vector(-3,2){0.5}}
\put(30,30){\line(-3,-4){15}}\put(21,18){\vector(-3,-4){0.5}}
\put(30,30){\line(-3,-2){15}}\put(21,24){\vector(-3,-2){0.5}}
\end{picture}}

\put(145,40){\begin{picture}(30,30) \put(30,30){\circle{2}}
\put(30,30){\circle{4}} \put(0,30){\circle{4}}
\put(0,30){\circle*{2}}\put(15,10){\circle*{3}}
\put(15,20){\circle*{3}}
\put(0,0){\line(1,0){30}}\put(0,0){\line(0,1){30}}
\put(30,0){\line(0,1){30}} \put(30,30){\line(-1,0){30}}
\put(15,0){\vector(-1,0){0.5}} \put(0,15){\vector(0,-1){0.5}}
\put(30,13){\vector(0,1){0.5}}\put(15,30){\vector(1,0){0.5}}
\put(0,0){\line(3,4){15}}\put(9,12){\vector(3,4){0.5}}
\put(30,0){\line(-3,4){15}}\put(21,12){\vector(3,-4){0.5}}
\put(0,0){\line(3,2){15}}\put(9,6){\vector(3,2){0.5}}
\put(30,0){\line(-3,2){15}}\put(21,6){\vector(3,-2){0.5}}
\put(0,30){\line(3,-4){15}}\put(9,18){\vector(-3,4){0.5}}
\put(0,30){\line(3,-2){15}}\put(9,24){\vector(-3,2){0.5}}
\put(30,30){\line(-3,-4){15}}\put(21,18){\vector(-3,-4){0.5}}
\put(30,30){\line(-3,-2){15}}\put(21,24){\vector(-3,-2){0.5}}
\end{picture}}

\put(145,25){\line(3,5){15}}\put(151,35){\vector(-1,-2){0.5}}
\put(175,25){\line(-3,5){15}}\put(169,35){\vector(-1,2){0.5}}

\put(145,40){\line(3,-5){15}}\put(151,30){\vector(1,-2){0.5}}
\put(175,40){\line(-3,-5){15}}\put(169,30){\vector(1,2){0.5}}

\put(130,47.5){\line(5,2){30}}\put(140,51.5){\vector(-2,-1){0.5}}
\put(160,60){\line(5,-2){30}}\put(170,56){\vector(-2,1){0.5}}

\put(130,17.5){\line(5,-2){30}}\put(140,13.5){\vector(2,-1){0.5}}
\put(160,5){\line(5,2){30}}\put(170,9){\vector(2,1){0.5}}

\put(97,32){\makebox(0,0){$0$}} \put(112,21){\makebox(0,0){$1$}}
\put(112,44){\makebox(0,0){$2$}} \put(226,32){\makebox(0,0){$21$}}

\put(211,44){\makebox(0,0){$22$}}
\put(211,21){\makebox(0,0){$23$}}
\put(198,32.5){\makebox(0,0){$18$}}
\put(122,32){\makebox(0,0){$3$}}

\put(138,39.5){\makebox(0,0){$8$}}
\put(127,13){\makebox(0,0){$5$}} \put(127,52){\makebox(0,0){$4$}}
\put(138,24.5){\makebox(0,0){$7$}}

\put(183,40){\makebox(0,0){$16$}}
\put(196,13){\makebox(0,0){$19$}}
\put(196,52){\makebox(0,0){$20$}}
\put(183,24.5){\makebox(0,0){$17$}}

\put(178,65){\makebox(0,0){$14$}} \put(178,0){\makebox(0,0){$13$}}
\put(141,65){\makebox(0,0){$10$}} \put(141,0){\makebox(0,0){$11$}}

\put(159.5,43){\makebox(0,0){$15$}}
\put(159,21){\makebox(0,0){$12$}} \put(159,65){\makebox(0,0){$9$}}
\put(160,0){\makebox(0,0){$6$}}

\put(100,60){\makebox(0,0){($\mathcal{E}_{21}$)}}

\end{picture}
\end{center}
\caption{Some graphs with self-fusion: The $\mathcal{A}_{k}$ series, $\mathcal{D}_9$, $\mathcal{E}_{5}$, $\mathcal{%
E}_{9}$ and $\mathcal{E}_{21}$.} 
\label{fig:graphself}
\end{figure}

The $\mathcal{D}_{9}^{t}$ graph\ is a module over the graph
algebra $\mathcal{A}_{9}$ and over the graph algebra $\mathcal{D}_{9}$. It is associated to the following
modular invariant partition function:
\begin{eqnarray*}
\mathcal{Z}(\mathcal{D}_9^{t}) &=&  |\chi_{(0,0)}^9+\chi_{(9,0)}^9+\chi_{(0,9)}^9|^2
+ |\chi_{(3,0)}^9+\chi_{(6,3)}^9+\chi_{(0,6)}^9|^2
+ |\chi_{(0,3)}^9+\chi_{(6,0)}^9+\chi_{(3,6)}^9|^2 \\
 &+&  |\chi_{(2,2)}^9+\chi_{(5,2)}^9+\chi_{(2,5)}^9|^2
+ |\chi_{(4,4)}^9+\chi_{(4,1)}^9+\chi_{(1,4)}^9|^2
+ 2\, |\chi_{(3,3)}^9|^2  \\
&+&  \left\lbrack(\chi_{(1,1)}^9+\chi_{(7,1)}^9+\chi_{(1,7)}^9)\,
\overline{\chi_{(3,3)}^9} + h.c. \right\rbrack
\end{eqnarray*}
The graph $\mathcal{D}_9^t$ appears as a module of its own algebra
of quantum symmetries (calculated from the modular splitting
equation). It is a generalization of the $E_7$ graph\footnote{The
$E_7$ graph should better be called $\mathcal{D}_{16}^t$.} of the
$SU(2)$ system. Its quantum mass is $m(\mathcal{D}_9^t) = 72(2+\sqrt{3})$. 
$Oc(\mathcal{D}_9^t)$ is obtained via an
anti-automorphism called the exceptional ambichiral twist $\xi$,
which acts on vertices of the modular subalgebra
$J=\{0_{0},2_{0},3_{0},3_{0}^{\prime
},4_{0},5_{0},\alpha_{0}^{1},\alpha _{0}^{2},\alpha _{0}^{3}\} $
of $\mathcal{D}_{9}$ (see Figure \ref{fig:graphself}), such
that $\xi \left( 2_{0}\right) =\alpha _{0}^{2},$ $\xi \left(
\alpha _{0}\right) =2_{0}$ and $\xi \left( u\right) =u$ for all
others $u\in J$. The Ocneanu algebra $Oc(\mathcal{D}_{9}^{t})$
involves $\mathcal{D}_{9}\otimesdot_{J}\mathcal{D}_{9}$ and a non
commutative matrix complement. We identify $a \otimesdot_{J} u \,
b \equiv a \xi(u^*) \otimesdot_{J} b$ for all $u\in J$ and $a,b\in
\mathcal{D}_{9}$. Its dimension is 55 and the quantum mass is $m(Oc(\mathcal{D}_9^t)) = m(\mathcal{A}_9)
= 432 (7+4\sqrt{3})$. The dimension is $\dim (\mathcal{BD}_{9}^{t}) = 1\,167\,355$.

\subsubsection{The twisted conjugate exceptional ${\mathcal{D}_{9}^{t}}^{\ast}$ (no self-fusion)}
The ${\mathcal{D}_{9}^{t}}^{\ast}$ graph is a module graph over the graph algebras $\mathcal{A}_9$, 
$\mathcal{D}_9$ and also $\mathcal{D}_9^t$. 
The modular invariant partition function associated to this graph is:
\begin{eqnarray*}
\mathcal{Z}({\mathcal{D}_{9}^{t}}^{*}) &=&  |\chi_{(0,0)}^9+\chi_{(9,0)}^9+\chi_{(0,9)}^9|^2
+ |\chi_{(2,2)}^9+\chi_{(5,2)}^9+\chi_{(2,5)}^9|^2
+ |\chi_{(4,4)}^9+\chi_{(4,1)}^9+\chi_{(1,4)}^9|^2 \\
 &+& 2\, |\chi_{(3,3)}^9|^2 +
\left\lbrack (\chi_{(0,3)}^9+\chi_{(6,0)}^9 + \chi_{(3,6)}^9) \,
(\overline{\chi_{(3,0)}^9} + \overline{\chi_{(6,3)}^9} + \overline{\chi_{(0,6)}^9}) + h.c.
\right\rbrack \\
&+& \left\lbrack (\chi_{(1,1)}^9+\chi_{(7,1)}^9+\chi_{(1,7)}^9)\,
\overline{\chi_{(3,3)}^9} + h.c. \right\rbrack
\end{eqnarray*}
The ${\mathcal{D}_{9}^{t}}^{*}$ graph appears as a module of its
own algebra of quantum symmetries, which is also obtained via the
exceptional ambichiral twist $\xi$ acting on vertices of $J
\subset \mathcal{D}_9$. The Ocneanu algebra
$Oc({\mathcal{D}_{9}^{t}}^{\ast})$ involves also
$\mathcal{D}_{9}\otimesdot_{J} \mathcal{D}_{9}$ and a non
commutative matrix complement, but with the identifications $a
\otimesdot_{J} u \, b = a \xi(u) \otimesdot_{J} b$ for all $u\in
J$ and $a,b\in \mathcal{D}_{9}$. Its dimension is 55 and the quantum mass is $m(Oc({\mathcal{D}_9^t}^*)) = m(Oc(\mathcal{D}_9^t)) = m(\mathcal{A}_9)$.  The dimension is
$\dim({\mathcal{BD}_{9}^{t}}^{\ast}) = 531\,435$.


\section{Comments}

\paragraph{Overall features of  quantum groupo\"{\i}ds and graphs associated to higher Coxeter-Dynkin systems}
For an $SU(n)$ system of graphs, one expects the  following pattern.  
The family of ${\cal A}_k$ graphs is easily obtained by
truncation of the Weyl chambers at level $k$;  such ${\mathcal
A}_k$ graphs involve several types of oriented lines (one for each
fundamental representation of $SU(n)$). 
Then one can obtain several
other families by using the existence of automorphisms such as
complex conjugacy (leading to the $\mathcal{A}_k^*$ series),
$\mathbb{Z}_p$ symmetries (leading to the orbifold $\mathcal{D}_k[p]
= \mathcal{A}_k/p$ series), or a combination of these two
automorphims (leading to the $\mathcal{D}_k^*[p]$ series). 
From our experience with small values of $n$, we expect rather different families of 
$\mathcal{D}$ graphs, depending on whether $n$ is even or odd. For $SU(2)$, orbifold graphs 
$\mathcal{D}_k[2] = \mathcal{D}_{\frac{k}{2}+2}$ exist if $k =0,2 \mod 4$, and they have self-fusion 
whenever $k=0\mod4$. 
For $SU(3)$, orbifold graphs $\mathcal{D}_k[3]$ exist for all $k$, and they have self-fusion whenever
$k=0\mod3$. For $SU(4)$, and according to \cite{Oc-Bariloche}, we have orbifold graphs of type 
$\mathcal{D}_k[2]$ for all $k$, and they have self-fusion whenever $k=0\mod 2$, but we have also 
orbifold graphs of type $\mathcal{D}_k[4]$ for $k=0,2,6 \mod 8$, and they have self-fusion whenever 
$k=0\mod8$. 

For $\mathcal{A}_k$ and $\mathcal{A}_k^*$ series, the algebra of
quantum symmetries can be determined from the tensor square
of the graph algebra $\mathcal{A}_k$, suitably quotiented.
When $\mathcal{D}_k$ does not have self-fusion, its algebra of
quantum symmetries can also be determined from the tensor square
of the graph algebra $\mathcal{A}_k$, suitably quotiented with the help of appropriate
generalizations of the Gannon twist. This is also the
case for its corresponding conjugated series.
When $\mathcal{D}_k$
graph has self-fusion, its algebra of quantum
symmetries (which, in this case, is non commutative) can be obtained as a
cross-product of the graph algebra of ${\mathcal D}_k$ by the
cyclic group $\mathbb{Z}_p$; this is also the case for the
corresponding conjugate series. In any of these cases, the
associated modular invariant is easy to obtain from the
$\mathcal{A}$ modular invariant at same level.

For a given system, it seems that one can always  find a (unique)
exceptional graph  ${\mathcal D}^t$, without self-fusion, whose
algebra of quantum symmetries is equal to the quotient  of the
tensor square of a particular ${\mathcal D}$ graph by an
exceptional automorphism (this generalizes the $(E_7,  D_{10})$
situation of the $SU(2)$ family). The graph  ${\mathcal D}^t$
itself is then recognized as a module over its algebra of quantum
symmetries. Determination of this automorphism can be found by
looking at the values of the modular operator $T$ on vertices of
the corresponding ${\mathcal A}$ graph and the
induction-restriction rules from ${\mathcal A}$ to ${\mathcal D}$
\cite{GilRobert2}. Same discussion for the corresponding
conjugated graph ${\mathcal{D}^t}^{*}$.

 We are then left with the other exceptional graphs. They may admit self-fusion or not. When they don't, they are
orbifolds of those exceptionals that enjoy self-fusion.  Graphs
with self-fusion are called ``quantum subgroups'' by A. Ocneanu,
the others being only ``quantum modules''. Those exceptional
subgroups are $E_6 \equiv {\mathcal E}_{10} $ and $E_8 \equiv
{\mathcal E}_{28}$ for the $SU(2)$ system,  ${\mathcal E}_5$,
${\mathcal E}_9$ and ${\mathcal E}_{21}$ for the $SU(3)$ system
and ${\mathcal E}_4$, ${\mathcal E}_6$ and ${\mathcal E}_8$ for
the $SU(4)$ system. Their algebra of quantum symmetries may be
commutative or not. Non commutativity can be deduced, either from
the presence of integer entries bigger than $1$ in the modular
invariant, or from the existence of non trivial classical
symmetries in the graph itself (see footnote in section 5.4).
When the algebra of quantum symmetries $Oc(G)$ is commutative, like for $E_6$ and $E_8$ in
the $SU(2)$ system, or like for ${\cal E}_5$, ${\cal E}_{21}$ in
the $SU(3)$ system, it is easy to obtain the corresponding toric
matrices and $Oc(G)$ itself without
having to solve the modular splitting equation, because, in these
cases, one obtains $Oc(G)$ as a tensor square of $G$ itself over
the modular subalgebra $J$ which can be determined by using  the
properties of the modular generator $T$ under
restriction-induction (see \cite{GilRobert2}). Of course, it is
always advisable to check that the obtained result satisfies the
modular splitting equation. If, however, the algebra of quantum symmetries of this
exceptional graph with self-fusion  is non commutative (like for
the ${\cal E}_{9}$ case), the determination of $Oc(G)$  becomes
quite involved and the only  method we can think of is again to
use the modular splitting technique.

Once the exceptional graphs with self-fusion are known, it is not
too difficult to obtain the exceptional modules : they are
quotients or orbifolds of the former and often appear as
particular subspaces of $Oc(G)$.

Finally, let us mention that when the graph $G$ is a priori known, and whenever 
the vertex $x$ of $Oc(G)$ can be written as $a \otimesdot b$, with $a,b \in G$, it is usually possible to obtain 
(or recover) the toric matrices $W_{x0}$ from the annular or essential matrices, see for instance
\cite{GilRobert1} or \cite{Gil-thesis}. This method, first presented in \cite{Coq-qtetra}, is 
particularly easy to implement when one considers generalizations of the exceptional graphs with 
self-fusion $E_6$ and $E_8$ (\ie $\mathcal{E}_5$ and $\mathcal{E}_{21}$ for the $SU(3)$ system), since 
$Oc(G)=G\otimesdot_J G$, in those cases. One obtains $W_{x0} = \sum_{c\in J} (F_{\lambda})_{ac}\,(F_{\lambda})_{bc}
= E_a . ((E_b)^{red})^T$, where the reduced essential matrices $E_b^{red}$ are obtained from the $E_b$ by 
keeping the matrix elements of those columns corresponding to the modular subalgebra $J$ and putting all others 
entries to zero. 

\paragraph{Graphs from modular invariants.}

One possibility is  to rely on a given classification of the
modular invariants. Such a classification exists for $SU(2)$
\cite{CIZ} and $SU(3)$ \cite{Gannon-su3} but is not available for
$SU(n)$ when $n>3$. However there are arguments showing that the
level of exceptionals cannot be too high \cite{Oc-MSRI}, so that
it is enough to explore a sizeable list of possibilities.  Once a
modular invariant is known,  one can  use the modular splitting
technique and find the algebra $Oc(G)$. Generically,  the Ocneanu
graph involves one or several copies of the graph $G$ itself and
of its modules; this may not be so in special cases, for instance
the $D_{odd}$ case of the $SU(2)$ system or in the conjugated
series of the $SU(3)$ system, but then, other techniques of
determination can be used (cf the above discussion). Once the
graph $G$ is obtained, one has still to check that the
obtained result gives rise to a ``good'' theory of representations
(here $SU(3)$); otherwise, it should be discarded. We believe that
the precise meaning of this sentence is that the obtained graph
should give rise to a Kuperberg spider \cite{KuperbergSpiders};
another possibility is to  use the existence of a
self-connection,  as defined by A. Ocneanu in
\cite{Oc-Bariloche}. As already mentioned, we believe that the two
notions coincide but it is clear that some more work is needed in
this direction. The list of graphs expected to provide an answer
to the $SU(4)$ classification problem is given in
\cite{Oc-Bariloche}.

\paragraph{Conformal embeddings}

 Another possibility leading to interesting candidates for  graphs $G$ of  higher Coxeter-Dynkin systems is to use the existence of  conformal embeddings of affine algebras -- a subject that we did not touch in this paper. One should be aware that  1) List of modular invariants,  2) List of conformal embeddings, 3) List of graphs belonging to higher Coxeter-Dynkin systems (or defining Ocneanu quantum groupo\"{\i}ds) are distinct problems.

 It happens that, for $SU(2)$ and $SU(3)$,  all exceptional graphs with self-fusion correspond to particular conformal embeddings, but other such embeddings lead   to  orbifolds or to   members (with small level) of the ${\mathcal D}$ series. In the case of $SU(4)$, it seems that there is one exceptional graph with self-fusion  not associated with any conformal embedding.

Conformal embeddings of affine algebras at level k  of the type $ \widehat{su}(n)_k  \subset \hat g_1 $,
where $g$ is a simple Lie algebra, simply laced or not, can be associated with graphs that are candidates to become members, at level $k$, of the Coxeter-Dynkin system of $SU(n)$.
The condition to be conformal imposes equality of the central charges :
\begin{equation}
\frac{(n^2 -1) k}{ {k + n}}  = \frac {dim(g)   }{1 + \kappa (g)}
\end{equation}
where $dim(g)$ is the dimension of $g$ and $\kappa (g)$ its {\sl
dual\/}  Coxeter number. This equation is easy to solve for all
$SU(n)$ systems.  In the case $n=2$ there are three non trivial
solutions: $E_6$ ($\equiv{ \mathcal E}_{10}$), for $g = B_2 =
spin(5) $, then  $E_8$( $\equiv { \mathcal E}_{28}$) for $g = G_2$
and finally $D_4$ ($\equiv { \mathcal D}_{4}$), for $g= A_2 =
su(3)$. In the case $n=3$ there are many more solutions; let us
just mention those that give rise to exceptionnal graphs with
self-fusion : ${\mathcal E}_{5}$ for $g=A_5 = su(6)$, then
${\mathcal E}_{9}$ for $g=A_6 = su(7)$ and finally   ${\mathcal
E}_{21}$ for $g=E_7$.

 \paragraph {Other generalizations.} The algebra of quantum symmetries  described in the previous section refers to quantum groupo\"{\i}ds for which a basis of matrix units, for the vertical product, is made of double triangles of type $GGAGG$, where $G$ is any graph of the system ($A$-type, $D$-type, exceptionnal type \etc). However one may replace these double triangles by others, of type $GGKGG$, whenever $G$ is a $K$ module. This was apparently not studied.

  \paragraph {About the definitions of $Oc(G)$}
 The most pleasant definition of $Oc(G)$ is to take it as the algebra of characters (or irreps) for the horizontal product on  $\widehat{\mathcal{B}}G$.  This amounts to consider the center of $\widehat{\mathcal{B}}G$ (for the horizontal multiplication $\widehat \circ$) and analyse its structure when endowed with a product  inherited from the vertical multiplication on ${\mathcal{B}G}$.
  However, to determine it in this way requires a priori the calculation
 of several sets (finite but huge) of generalized $6J$ symbols. It seems that nobody ever did it this way (the family of $6J$ symbols is not even known for the exceptional cases of the $SU(2)$ system !). Rather, the generators $O_x$ were obtained as explained in step II of the modular splitting technique.  A  clear discussion relating these two types of concepts would be welcome.

 \paragraph{Frontiers.}
The possibility of associating higher order algebraic systems
(somehow generalizing universal envelopping algebras and their
root systems) to graphs that are members of higher Coxeter-Dynkin
families is certainly a fascinating perspective, which was not
discussed in this paper.

 \paragraph{Conclusion.}
The quantum groupo\"{\i}d aspects of these systems are still
largely under-studied. As already stated previously,  and in
agreement with popular wisdom, every graph $G$ belonging to an
$SU(n)$ system should give rise, and conversely,  to an ``Ocneanu
quantum groupo\"{\i}d''. All together these objects constitute a
particular family of finite dimensional weak Hopf algebras.
However, many general properties still need clarification and
every single particular diagram should deserve more study -- for
instance the explicit determination of the different types of
cells (generalized $6J$ symbols), is an open problem.

\vspace{1.0cm}

\noindent {\bf \Large Acknowledgements}
D.H. and E.H.T. are grateful to FRUNAM and CIRM (Marseille) for supporting their stay at CIRM where part of this work was done. R. C. is grateful to the pole RENAPT to support his stay at LPTP in Oujda. G.S. would like to thank AUF -- Agence Universitaire de la Francophonie  -- for financial support.


\end{document}